\documentclass[lettersize,journal]{IEEEtran}
\usepackage{amsmath,amsfonts}
\usepackage[hidelinks]{hyperref}
\usepackage{algorithmic}

\usepackage{algorithm}
\usepackage{array}
\usepackage[caption=false,font=normalsize,labelfont=sf,textfont=sf]{subfig}
\usepackage{textcomp}
\usepackage{stfloats}
\usepackage{url}
\usepackage{verbatim}
\usepackage{graphicx}
\usepackage{cite}
\usepackage{amssymb}
\usepackage{bm}
\usepackage[table]{xcolor}
\usepackage{colortbl}
\usepackage{array}  
\usepackage{pifont}  
\usepackage{multirow}
\usepackage{caption}
\newcommand{\panel}[1]{%
  \includegraphics[width=0.245\textwidth,trim=30mm 85mm 30mm 85mm,clip]{#1}%
}

\newlength{\AlgStageBefore}
\newlength{\AlgStageAfter}

\begin{document}

\title{Diffusion-Based Noise-Adaptive Null-Space Channel Estimation for OFDM Systems}

\author{Heqiang Qi, Yirun Chen, Xiangming Meng$^{\dagger}$, Chunxiao Jiang,~\IEEEmembership{Fellow,~IEEE,}\\ Sheng Wu,~\IEEEmembership{Member,~IEEE,} and Linling Kuang,~\IEEEmembership{Member,~IEEE}
\thanks{Heqiang Qi is with the College of Information Science and Electronic Engineering, Zhejiang University, Hangzhou 310027, China (e-mail: \href{heqiangqi@zju.edu.cn}{heqiangqi@zju.edu.cn}).}
\thanks{Yirun Chen and Xiangming Meng are with Zhejiang University-University of Illinois Urbana--Champaign Institute, Zhejiang University, Haining 314400, China (e-mail: \href{mailto:yirun.25@intl.zju.edu.cn}{{yirun.25@intl.zju.edu.cn}}; \href{mailto:xiangmingmeng@intl.zju.edu.cn}{{xiangmingmeng@intl.zju.edu.cn}}).}
\thanks{Chunxiao Jiang and Linling Kuang are with the Beijing National Research Center for Information Science and Technology, and the State Key Laboratory of Space Network and Communications, Tsinghua University, Beijing 100084, China (e-mail: \href{mailto:jchx@tsinghua.edu.cn}{jchx@tsinghua.edu.cn}; 
\href{mailto:kll@tsinghua.edu.cn}{kll@tsinghua.edu.cn}).}
\thanks{Sheng Wu is with the School of
Information and Communication Engineering, Beijing University of Posts
and Telecommunications, Beijing 100876, China (e-mail: \href{thuraya@bupt.edu.cn}{thuraya@bupt.edu.cn}).}
\thanks{$^{\dagger}$Corresponding author: Xiangming Meng.}
}



\maketitle

\begin{abstract}

Accurate channel estimation in orthogonal frequency division multiplexing (OFDM) systems remains challenging when demodulation reference signal (DMRS) observations are sparse and noisy, and when DMRS configurations vary across deployment scenarios. This paper proposes DANCE (Diffusion-based Noise-Adaptive Null-space Channel Estimation), a diffusion-based channel estimator for OFDM systems. We formulate DMRS-aided channel estimation as a sparse linear inverse problem whose measurement operator is induced by the pilot pattern. 
The resulting range--null space decomposition separates the measurement-constrained range-space component from the unobserved null-space component, which is reconstructed through a learned diffusion prior.
To avoid directly imposing noisy pilot samples as exact constraints, DANCE introduces a noise-adaptive posterior correction into the reverse diffusion process. The correction coefficient and the residual sampling variance are jointly calibrated according to the observation noise level, thereby reducing pilot-noise injection while retaining useful measurement information. 
We further design a conditional U-Net denoiser for complex-valued OFDM channel grids, where the real and imaginary components are represented as separate feature channels and downsampling is performed only along the subcarrier dimension. Simulations based on 5G NR tapped delay line (TDL) and clustered delay line (CDL) channel models show that DANCE achieves consistently lower normalized mean squared error (NMSE) than conventional estimators and diffusion-based posterior sampling methods under different signal-to-noise ratios, DMRS configurations, Doppler frequency shifts, and train--test distribution mismatches.

\end{abstract}


\section{Introduction}

\IEEEPARstart{W}{ith} the continued deployment of fifth-generation (5G) networks and the development of sixth-generation (6G) wireless systems, radio links are expected to support high data rates, reliable connectivity, and user mobility in increasingly diverse propagation environments~\cite{andrews2014will,saad2019vision,zhang20196g}. Orthogonal frequency division multiplexing (OFDM) remains a central multicarrier technique because it converts a frequency-selective channel into parallel narrowband subchannels and enables efficient equalization~\cite{li2006orthogonal}. At the receiver, coherent demodulation and data detection depend on accurate channel state information (CSI) over the time--frequency resource grid. Reliable channel estimation is therefore central to OFDM receiver design and has a direct impact on link reliability and throughput.

In practical OFDM systems, CSI is commonly inferred from demodulation reference signals (DMRSs) placed on a subset of resource elements (REs)~\cite{channels2020nr}. The receiver observes the channel only at these pilot REs and must infer the response at both pilot and data REs. This reconstruction is difficult for three related reasons. First, the pilot density is deliberately limited to preserve spectral efficiency, so most entries of the channel grid are unobserved. Second, the DMRS pattern can change with the selected configuration and deployment requirement, which changes the measurement operator seen by the estimator. Third, the available pilot observations are noisy, and the channel may vary rapidly in high-mobility conditions. From this viewpoint, DMRS-aided OFDM channel estimation can be regarded as a sparse and noisy linear inverse problem~\cite{stuart2010inverse} with a configuration-dependent measurement pattern.

Classical estimators provide important baselines for this problem. Least squares (LS) estimation is simple and requires little prior information, but the pilot-domain estimate is sensitive to noise and the subsequent interpolation to non-pilot REs becomes unreliable when pilot density is low. Linear minimum mean-square error (LMMSE) estimation can improve accuracy by using second-order channel statistics, but its performance depends on accurate covariance and noise-variance information, which may be unavailable or mismatched in time-varying deployments~\cite{kay1993fundamentals}. These limitations become more pronounced when sparse pilots, changing DMRS configurations, and mobility-induced channel variation occur simultaneously.

Driven by advances in deep learning~\cite{lecun2015deep,he2016deep,krizhevsky2012imagenet} and its increasing integration with wireless communications~\cite{dai2020deep,shen2023five}, learning-based channel estimators have been developed to reduce the dependence on hand-crafted interpolation and explicit statistical models. Multilayer perceptrons have been used for end-to-end OFDM receiver processing or for refining coarse channel estimates~\cite{ye2017power,le2021deep}. Convolutional neural networks exploit local time--frequency correlations for CSI interpolation and denoising~\cite{soltani2019deep,li2019deep}, recurrent architectures model temporal channel evolution in doubly selective environments~\cite{gizzini2023rnn}, and attention-based networks, inspired by the Transformer architecture~\cite{vaswani2017attention}, capture longer-range interactions over the resource grid for wireless channel estimation~\cite{luan2022attention}. These methods have shown strong empirical performance, but they typically learn a direct mapping from a particular form of observation or coarse estimate to the full channel grid. As a consequence, their behavior can be sensitive to changes in pilot pattern, channel statistics, or deployment condition relative to the training data.

Generative models offer a complementary route. Rather than learning only an observation-to-channel mapping, a generative model attempts to learn the distribution of plausible channel responses, which can then be combined with measurement constraints for inverse reconstruction. Earlier generative approaches based on variational autoencoders (VAEs)~\cite{kingma2013auto} and generative adversarial networks (GANs)~\cite{goodfellow2020generative} have motivated high-dimensional channel estimation and wireless signal recovery with learned priors~\cite{balevi2020high,fesl2024channel}. In practice, however, VAEs may have limited representation capability for complex high-dimensional distributions, whereas GANs can be sensitive to training instability and architectural choices~\cite{huang2018introvae,kurach2019large}. Diffusion models are attractive in this setting because their training objective is stable and their iterative denoising process provides a natural mechanism for representing structured uncertainty~\cite{sohl2015deep,ho2020denoising,song2020score,dhariwal2021diffusion}.

Recent work has begun to apply diffusion models to wireless channel estimation, confirming that score-based or diffusion priors can improve channel recovery in high-dimensional settings~\cite{arvinte2022mimo,zhou2025generative}. Nevertheless, diffusion-based studies explicitly tailored to OFDM channel estimation remain limited, especially for DMRS-aided reconstruction over a sparse time--frequency resource grid. At the same time, generic diffusion-based inverse problem solvers provide useful mechanisms for incorporating measurements during sampling~\cite{daras2024survey}, including diffusion posterior sampling (DPS)~\cite{chung2022diffusion}, diffusion model based posterior sampling (DMPS)~\cite{meng2022diffusion} for noisy linear inverse problems, and denoising diffusion null-space models (DDNM)~\cite{wang2022zero} for observation-constrained generation. Taken together, existing diffusion-based methods provide valuable algorithmic tools, but they do not explicitly address the structural characteristics of DMRS-based OFDM channel estimation: the measurement matrix is sparse and diagonal, the observed entries are corrupted by pilot noise, the pilot pattern may vary across configurations, and the learned prior may be tested under channel distributions different from those used for training.

This paper proposes DANCE (\textbf{D}iffusion-based Noise-\textbf{A}daptive \textbf{N}ull-space \textbf{C}hannel \textbf{E}stimation), a diffusion-based channel estimator explicitly tailored to OFDM channel estimation with sparse DMRS observations. We formulate the DMRS-aided estimation task as a linear inverse problem whose measurement operator is determined by the DMRS pattern and is sparse and diagonal. DANCE then uses the range--null space decomposition induced by this operator to separate the measurement-constrained range-space component from the unobserved null-space component. The former is corrected using the measured DMRS samples, whereas the latter is recovered through a learned diffusion prior. Since the DMRS observations are noisy, DANCE does not impose them as exact constraints. Instead, it introduces a noise-adaptive posterior correction that adjusts both the range-space correction strength and the residual sampling variance in the reverse diffusion process according to the observation noise level. This allows the reverse diffusion process to exploit pilot information while limiting pilot-noise injection into the reconstructed full-grid channel. In addition, DANCE employs an OFDM-tailored conditional U-Net denoiser for complex-valued channel grids, where the real and imaginary components are represented as separate feature channels and downsampling is performed only along the subcarrier dimension to preserve the OFDM symbol structure. Channel-scenario conditioning is further incorporated through classifier-free guidance (CFG)~\cite{ho2022classifier}  during reverse sampling, steering the denoising process toward the target channel distribution.

The main contributions of this paper are summarized as follows:
\begin{itemize}
\item We formulate DMRS-aided OFDM channel estimation as a sparse linear inverse problem and develop DANCE, a diffusion-based estimator that combines range--null space decomposition with a learned channel prior. The formulation explicitly separates the pilot-observed range-space component from the unobserved null-space component, providing a principled way to reconstruct full-grid CSI from sparse DMRS observations.

\item We introduce a noise-adaptive posterior correction mechanism for noisy pilot observations. The correction coefficient and the residual sampling variance are jointly determined from the observation noise level, allowing the reverse diffusion process to balance pilot consistency and noise suppression.

\item We design a conditional U-Net denoiser for complex-valued OFDM channel grids. The network represents the real and imaginary parts as separate feature channels, incorporates channel-scenario conditioning through classifier-free guidance, and performs downsampling only along the subcarrier dimension to preserve the OFDM symbol structure.

\item We evaluate DANCE using 5G NR TDL and CDL channel models under different signal-to-noise ratios (SNRs), DMRS configurations, Doppler frequency shifts, and train--test distribution mismatches. The results show that DANCE achieves lower normalized mean squared error (NMSE) than conventional estimators and diffusion-based posterior sampling baselines in the considered sparse-pilot settings.

\end{itemize}

\textbf{Notations:} For any matrix $\mathbf{A}$, $\mathbf{A}^{\mathrm{T}}$, $\mathbf{A}^{\mathrm{H}}$, and $\mathbf{A}^{\dagger}$ represent the transpose, conjugate transpose, and pseudo-inverse of $\mathbf{A}$, respectively. Also, $\mathbf{0}$ is a zero vector, $\mathbf{I}$ is an identity matrix, $\lVert \cdot \rVert_2$ denotes the $\ell_2$-norm, $\odot$ denotes the Hadamard product, and $\mathbb{E}[\cdot]$ denotes the expectation operator. $\mathcal{N}(z; \mu, \sigma^2)$ represents a Gaussian random variable $z$ with mean $\mu$ and variance $\sigma^2$. Let $\mathbb{C}$ and $\mathbb{R}$ represent the sets of complex and real numbers, respectively.  Finally, $[N]$ stands for the set of integers from 1 to $N$, and $\mathcal{U}([N])$ indicates a discrete uniform distribution over this set.

\section{Preliminaries}
\subsection{OFDM Channel Estimation}
In the 5G NR standard, the DMRS serves as the reference signal for data demodulation. Its core function is to enable the receiver to estimate the CSI in real time and accurately, thereby facilitating the correct recovery of the original transmitted data from the distorted received signal.

We adhere to the typical configuration for OFDM systems as defined by the 3GPP standards~\cite{channels2020nr}. The minimum unit of resource in the time domain is one OFDM symbol, and in the frequency domain, it is one subcarrier. The smallest time--frequency resource unit is defined as an RE, occupying one OFDM symbol in the time domain and one subcarrier in the frequency domain. Pilot symbols are sparsely placed on a subset of these REs, with their locations known a priori to both the transmitter and the receiver. All remaining REs within the allocated Resource Blocks (RBs) are dedicated to data transmission. The receiver uses these DMRS observations to reconstruct the full-grid CSI over both pilot and data REs.

\begin{figure}[!t]
    \centering
    \includegraphics[width=\columnwidth]{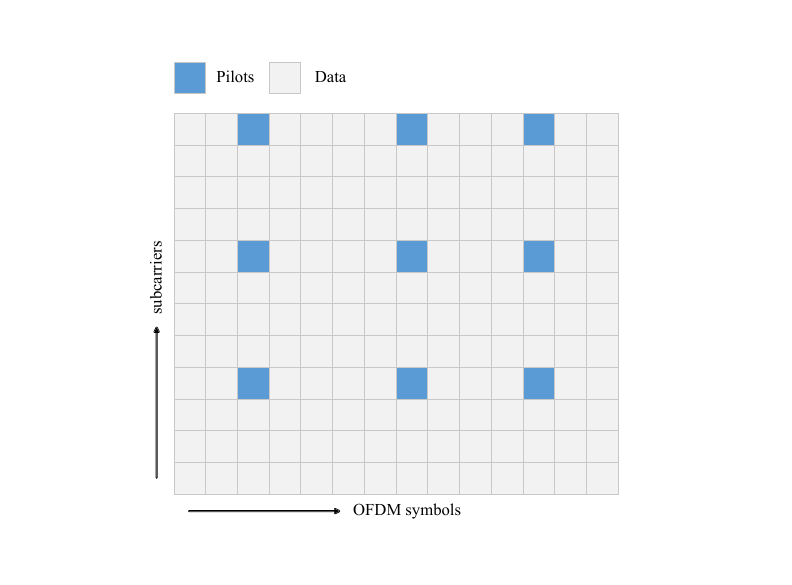}
    \caption{A representative example of a DMRS configuration. The REs highlighted in blue denote DMRS allocations, while the remaining REs are used for data transmission.}
    \label{fig:DMRS}
\end{figure}

5G NR supports multiple DMRS configurations. These configurations determine the pilot position in both the time and frequency domains, allowing the system to better adapt to various communication scenarios and requirements. A representative example of a DMRS configuration is illustrated in Fig.~\ref{fig:DMRS}.

The channel estimation problem for OFDM systems can be abstractly formulated as:
\begin{equation}
\label{OFDM_equation}
    \mathbf{Y} = \mathbf{P} \odot \mathbf{H} + \mathbf{N},
\end{equation}
where $\mathbf{H} \in \mathbb{C}^{N \times M}$ denotes the channel time--frequency resource blocks in a practical scenario, $\mathbf{P} \in \mathbb{C}^{N \times M}$ denotes the sparse DMRS pilot-symbol grid whose support is determined by the DMRS pattern, $\mathbf{Y} \in \mathbb{C}^{N \times M}$ is the received signal at the receiver, and $\mathbf{N} \in \mathbb{C}^{N \times M}$ denotes additive white Gaussian noise (AWGN). Here, $N$ and $M$ denote the numbers of subcarriers and OFDM symbols in the considered resource grid, respectively. The objective of channel estimation is to reconstruct the true unknown high-dimensional channel $\mathbf{H}$ from the known observation $\mathbf{Y}$.

Through a simple matrix vectorization transformation, the problem can be fully converted into a matrix multiplication form:
\begin{equation}
\label{OFDM_inverse_equation}
    \mathbf{y} = \mathbf{A} \mathbf{h} + \mathbf{n},
\end{equation}
where $\mathbf{y} \in \mathbb{C}^{NM \times 1}$, $\mathbf{A} \in \mathbb{C}^{NM \times NM}$, $\mathbf{h} \in \mathbb{C}^{NM \times 1}$, and $\mathbf{n} \in \mathbb{C}^{NM \times 1}$. Consequently, the channel estimation task can be regarded as a standard linear inverse problem~\cite{stuart2010inverse,daras2024survey}.

\subsection{Diffusion Models}
DMs~\cite{sohl2015deep,ho2020denoising,song2020score,dhariwal2021diffusion} are one of the most advanced and representative generative models. They consist of two distinct processes: the forward (a.k.a noising) process and the reverse (a.k.a denoising) process. In the forward process, noise is progressively added to the original clean data samples until they are transformed into pure Gaussian noise with zero mean. Conversely, the reverse process aims to gradually denoise the Gaussian noise, recovering the underlying data distribution, and thereby generating high-quality samples.

Specifically, given a clean data sample $\mathbf{h}_0 \sim q(\cdot)$, the forward process will generate a sequence of noisy data, i.e., $\mathbf{h}_1, \mathbf{h}_2, ..., \mathbf{h}_T$, where $T$ is the total number of steps. This process can be formulated using a Gaussian kernel,
\begin{equation}
    q(\mathbf{h}_t|\mathbf{h}_{t-1}) = \mathcal{N}(\mathbf{h}_t;\sqrt{1-\beta_t}\mathbf{h}_{t-1},\beta_t \mathbf{I}),
\end{equation}
where $\{\beta_t \in (0,1)\}_{t=1}^T$ represents the noise schedule of the forward diffusion process, and satisfies $0<\beta_1<\beta_2<...<\beta_T<1$. With the progressive addition of noise, the original data will gradually lose its structure information and as $T \rightarrow \infty $, $\mathbf{h}_T$ will converge to an isotropic Gaussian distribution. Defining $\alpha_t \triangleq 1-\beta_t$, $\bar{\alpha}_t \triangleq \prod_{i=1}^t \alpha_i$, it is straightforward to derive that
\begin{equation}
    \mathbf{h}_t = \sqrt{\bar{\alpha}_t} \mathbf{h}_0 + \sqrt{1-\bar{\alpha}_t} \mathbf{\epsilon}_t, \quad \mathbf{\epsilon}_t \sim \mathcal{N}(\mathbf{0},\mathbf{I}).
\end{equation}
So, the forward process probability model can be formulated as
\begin{equation}
    q(\mathbf{h}_t|\mathbf{h}_0) = \mathcal{N}(\mathbf{h}_t;\sqrt{\bar{\alpha}_t}\mathbf{h}_0, (1-\bar{\alpha}_t)\mathbf{I}).
\end{equation}

The reverse process is in fact the inversion of the forward process. Based on the availability of $q(\mathbf{h}_{t-1}|\mathbf{h}_t)$, an initial Gaussian noise can eventually be transformed into the true data. However, the calculation of $q(\mathbf{h}_{t-1}|\mathbf{h}_{t})$ is intractable. When this conditional probability is further conditioned on $\mathbf{h}_0$, it becomes a Gaussian distribution
\begin{equation}
    q(\mathbf{h}_{t-1}|\mathbf{h}_t, \mathbf{h}_0) = \mathcal{N}(\mathbf{h}_{t-1}; {\tilde{\mathbf{\mu}}}(\mathbf{h}_t, \mathbf{h}_0), \tilde{\beta}_t \mathbf{I}),
\end{equation}
and we can derive that
\begin{align}
    \label{mean}
    {\tilde{\mathbf{\mu}}}(\mathbf{h}_t, \mathbf{h}_0) &= \frac{1}{\sqrt{\alpha_t}}(\mathbf{h}_t - \frac{1-\alpha_t}{\sqrt{1-\bar{\alpha}_t}}\mathbf{\epsilon}_t), \\
    \tilde{\beta}_t &= \frac{1-\bar{\alpha}_{t-1}}{1-\bar{\alpha}_t}\beta_t,
\end{align}
where $\mathbf{\epsilon}_t$ denotes the noise perturbation of the sample at time step $t$. We then can use a neural network to parameterize this distribution
\begin{equation}
    p_\theta(\mathbf{h}_{t-1}|\mathbf{h}_t, \mathbf{h}_0) = \mathcal{N}(\mathbf{h}_{t-1};\mathbf{\mu}_\theta(\mathbf{h}_t, t), \tilde{\beta}_t \mathbf{I}).
    \label{postrior sampling}
\end{equation}
In other words, we need to train $\mathbf{\mu}_\theta(\mathbf{h}_t,t)$ to approximate the true Gaussian mean 
$\tilde{\mathbf{\mu}}(\mathbf{h}_t,\mathbf{h}_0)$. Since~\eqref{mean} shows that this mean can be equivalently expressed through the perturbation noise $\mathbf{\epsilon}_t$, we employ a denoising neural network $\mathbf{\epsilon}_\theta(\mathbf{h}_t,t)$ to estimate this noise component. The training objective of the diffusion model is written as
\begin{equation}
    \mathcal{L}_{DM}(\theta) =
    \mathbb{E}_{t\sim\mathcal{U}[T], \mathbf{h}_0, \mathbf{\epsilon}_t}
    \left[
    \left\|
    \mathbf{\epsilon}_t -
    \mathbf{\epsilon}_\theta(\mathbf{h}_t,t)
    \right\|_2^2
    \right],
\end{equation}
where $t\sim\mathcal{U}[T]$ denotes that the diffusion timestep is uniformly sampled from all $T$ steps, and the objective corresponds to the mean squared error between the injected noise $\mathbf{\epsilon}_t$ and the network-predicted noise $\mathbf{\epsilon}_\theta(\mathbf{h}_t,t)$.

After model training, the iterative sampling process can be expressed as
\begin{equation}
    \mathbf{h}_{t-1} = \frac{1}{\sqrt{\alpha_t}}(\mathbf{h}_t - \frac{1-\alpha_t}{\sqrt{1-\bar{\alpha}_t}}\mathbf{\epsilon}_\theta(\mathbf{h}_t, t)) + \sigma_t \mathbf{z}, \quad \mathbf{z}\sim \mathcal{N}(\mathbf{0}, \mathbf{I}),
\label{reverse_equation_diffusion}
\end{equation}
where $\sigma_t$ denotes the injection of random noise during the generation process to model the multimodality of the data distribution. Furthermore, it is also demonstrated in~\cite{ho2020denoising} that $\sigma_t^2 = \beta_t$ and $\sigma_t^2 = \tilde{\beta}_t$ yield experimentally equivalent performance. In our work, the former configuration is adopted.
In addition, it is established that the noise estimate $\mathbf{\epsilon}_\theta(\mathbf{h}_t, t)$ at a given sample point $\mathbf{h}_t$ is directly proportional to the score function $\nabla_{\mathbf{h}_t} \log p(\mathbf{h}_t)$ of its prior distribution. The precise relationship is given by~\cite{kingma2021variational}:
\begin{equation}
    \nabla_{\mathbf{h}_t} \log p(\mathbf{h}_t) = -\frac{\mathbf{\epsilon}_\theta(\mathbf{h}_t, t)}{\sqrt{1 - \bar{\alpha}_t}}.
\end{equation}
Consequently, the reverse denoising process of the diffusion model, which corresponds to the sample generation procedure, can be formally expressed by the following stochastic update rule:
\begin{equation}
    \begin{split}
       \mathbf{h}_{t-1} = \frac{1}{\sqrt{\alpha_t}} (\mathbf{h}_t + ({1 - \alpha_t}) \, \nabla_{\mathbf{h}_t} \log p(\mathbf{h}_t)) + \sigma_t \mathbf{z}, \\
       \quad \mathbf{z} \sim \mathcal{N}(0, I).
    \end{split}
\end{equation}
When $t=0$, the denoising process completes and $\mathbf{h}_0$ becomes the final sample drawn from the learned prior distribution.

\section{Method}
\begin{figure}[!t]
    \centering
    \includegraphics[width=\columnwidth]{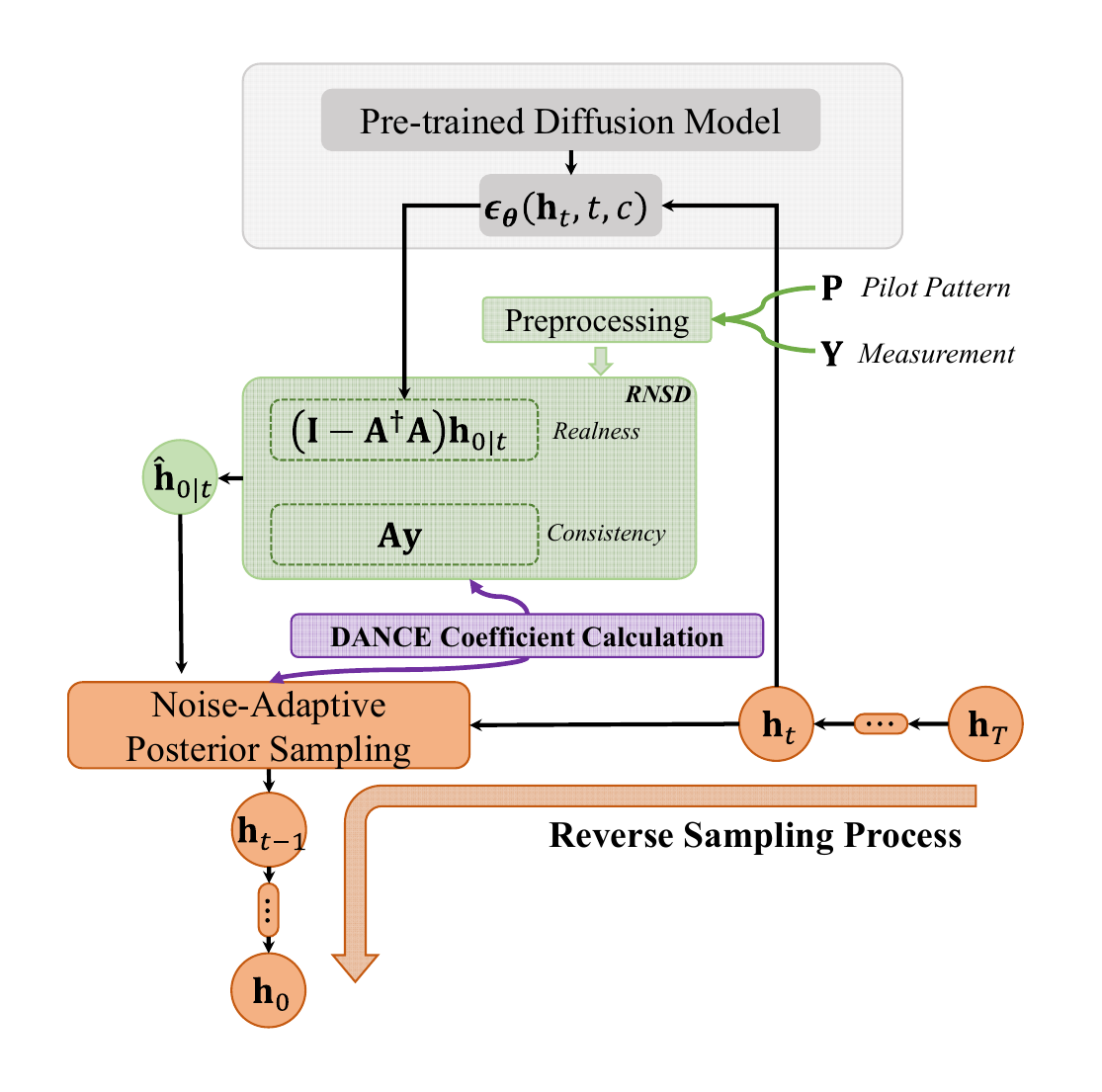}
    \caption{Block diagram of our proposed method DANCE.}
    \label{fig:system}
\end{figure}

This section presents the proposed DANCE framework. We first describe the range--null space decomposition induced by the DMRS measurement operator and then show how it is combined with a diffusion prior for OFDM channel estimation. The noise-adaptive posterior correction and the denoising network architecture are subsequently introduced. The overall framework is illustrated in Fig.~\ref{fig:system}.

\subsection{Range--Null Space Decomposition}
For a given matrix $\mathbf{A}\in \mathbb{C}^{M \times N}$, its pseudo-inverse, denoted $\mathbf{A}^{\dagger} \in \mathbb{C}^{N \times M}$, satisfies the property $\mathbf{A}\mathbf{A}^{\dagger}\mathbf{A} = \mathbf{A}$. From this property, the product $\mathbf{A}^{\dagger}\mathbf{A}$ defines a projection operator. Specifically, for any vector $\mathbf{h} \in \mathbb{C}^{N\times1}$, the operation $\mathbf{A}^{\dagger}\mathbf{A}\mathbf{h}$ yields the orthogonal projection of $\mathbf{h}$ onto the range space of $\mathbf{A}$, which is verified by the relation $\mathbf{A}\mathbf{A}^{\dagger}\mathbf{A}\mathbf{h} = \mathbf{A}\mathbf{h}$. In contrast, the complementary operator $(\mathbf{I} - \mathbf{A}^{\dagger}\mathbf{A})$ projects a vector onto the null space of $\mathbf{A}$, as it satisfies $\mathbf{A}(\mathbf{I} - \mathbf{A}^{\dagger}\mathbf{A})\mathbf{h} = \mathbf{0}$ for any $\mathbf{h}$.

Consequently, any vector $\mathbf{h}$ admits a unique orthogonal decomposition governed by these two projection operators: one component resides in the range space of $\mathbf{A}$, and the other lies in its null space. This decomposition is formally expressed as:
\begin{equation}
    \mathbf{h} = \mathbf{A}^{\dagger}\mathbf{A}\mathbf{h} + (\mathbf{I} - \mathbf{A}^{\dagger}\mathbf{A})\mathbf{h}.
\end{equation}

For the DMRS-based OFDM model considered in this paper, $\mathbf{A}$ is a highly sparse diagonal matrix determined by the pilot pattern, with nonzero diagonal entries only at the DMRS positions and zeros elsewhere. The above decomposition therefore has a direct interpretation: the range-space component corresponds to the pilot-observed entries, while the null-space component corresponds to the unobserved REs.

\subsection{Diffusion Model-Based Channel Estimation}
Building upon the foundation of range--null space decomposition and a key insight from~\cite{wang2022zero}, we specifically begin our analysis with the noiseless channel reconstruction problem, which is formulated as
\begin{equation}
    \mathbf{y} = \mathbf{A} \mathbf{h},
\end{equation}
where $\mathbf{h} \in \mathbb{C}^{NM \times 1}$ denotes the vectorized true channel, $\mathbf{A} \in \mathbb{C}^{NM \times NM}$ is the measurement matrix defined by the pilot pattern, and $\mathbf{y} \in \mathbb{C}^{NM \times 1}$ represents the noiseless observation. Given the input $\mathbf{y}$, the core objective of this inverse problem is to obtain an estimate $\hat{\mathbf{h}} \in \mathbb{C}^{NM \times 1}$ that satisfies two constraints: (i) \emph{Consistency}: $\mathbf{A} \hat{\mathbf{h}} = \mathbf{y}$; and (ii) \emph{Realness}: $\hat{\mathbf{h}} \sim q(\mathbf{h})$, where $q(\mathbf{h})$ denotes the distribution of the underlying  channels.

For the consistency constraint, we leverage the range--null space decomposition. As previously stated, the true channel $\mathbf{h}$ can be decomposed into its range-space component $\mathbf{A}^{\dagger} \mathbf{A} \mathbf{h}$ and its null-space component $(\mathbf{I} - \mathbf{A}^{\dagger} \mathbf{A}) \mathbf{h}$. The range-space component directly contributes to the observation $\mathbf{y}$, while the null-space component, by definition, lies in the null space of $\mathbf{A}$ and thus produces $\mathbf{0}$:
\begin{equation}
    \mathbf{A} \mathbf{h} = \mathbf{A} \mathbf{A}^{\dagger} \mathbf{A} \mathbf{h} + \mathbf{A} (\mathbf{I} - \mathbf{A}^{\dagger} \mathbf{A}) \mathbf{h} = \mathbf{A} \mathbf{h} + \mathbf{0} = \mathbf{y}.
\end{equation}

Given the observation $\mathbf{y}$, a general solution satisfying $\mathbf{A} \hat{\mathbf{h}} = \mathbf{y}$ can be directly constructed as
\begin{equation}
    \hat{\mathbf{h}} = \mathbf{A}^{\dagger} \mathbf{y} + (\mathbf{I} - \mathbf{A}^{\dagger} \mathbf{A}) \tilde{\mathbf{h}},
\label{equation_1}
\end{equation}
where $\tilde{\mathbf{h}}$ is an arbitrary vector. While any choice of $\tilde{\mathbf{h}}$ preserves consistency, it determines whether $\hat{\mathbf{h}}$ conforms to the true distribution $q(\mathbf{h})$. To this end, we employ a diffusion model to generate the null-space component $(\mathbf{I} - \mathbf{A}^{\dagger} \mathbf{A}) \tilde{\mathbf{h}}$, ensuring $\tilde{\mathbf{h}} \sim q(\mathbf{h})$. The core strength of such generative models lies in their ability to learn the complex prior distribution $q(\mathbf{h})$ of the channel from data. Consequently, the reconstructed channel satisfies both the data-consistency constraint and the distributional plausibility imposed by the learned prior.

Integrating with the formulation of diffusion models, an estimate of the initial channel $\mathbf{h}_0$ at timestep $t$ can be derived via Tweedie's formula~\cite{ho2020denoising}:
\begin{equation}
    \mathbf{h}_{0|t} = \frac{1}{\sqrt{\bar{\alpha}_t}}(\mathbf{h}_t - {\sqrt{1-\bar{\alpha}}_t}\mathbf{\epsilon}_\theta(\mathbf{h}_t, t)).
\label{equation_2}
\end{equation}
By choosing the arbitrary vector $\tilde{\mathbf{h}}$ in the general solution of~\eqref{equation_1} as the diffusion-based estimate $\mathbf{h}_{0|t}$ in~\eqref{equation_2}, the reconstructed channel can be written as:
\begin{equation}
    \hat{\mathbf{h}}_{0|t} = \mathbf{A}^{\dagger} \mathbf{y} + (\mathbf{I} - \mathbf{A}^{\dagger} \mathbf{A}) \mathbf{h}_{0|t}.
    \label{noiseless equation}
\end{equation}
Given the current state $\mathbf{h}_t$, the consistency constraint provided by observation $\mathbf{y}$, and the realism prior learned by the diffusion model, the state $\mathbf{h}_{t-1}$ can be updated according to~\eqref{postrior sampling}. This gives the noiseless range--null space reconstruction form, which serves as the basis for the noise-adaptive posterior correction introduced in the next subsection. 

\subsection{Noise-Adaptive Posterior Correction}
\label{subsec:noise_adaptive_calibration}

In practical OFDM channel estimation, the received pilot observations are modeled as noisy linear measurements of the underlying channel response:
\begin{equation}
    \mathbf{y} = \mathbf{A} \mathbf{h} + \mathbf{n}, 
    \qquad 
    \mathbf{n} \sim \mathcal{N}(\mathbf{0},\sigma_y^2 \mathbf{I}),
    \label{eq:noisy_observation}
\end{equation}
where $\sigma_y^2$ denotes the variance of the additive white Gaussian noise. Unlike the noiseless case, the observation $\mathbf{y}$ cannot be treated as an exact linear constraint on the channel. If the range-space component is directly enforced according to the noisy observation, the noise term in $\mathbf{y}$ may also be projected into the reconstructed channel, leading to performance degradation, especially in the low-SNR regime. Therefore, the observation consistency should be imposed in a noise-aware manner. To this end, we formulate the reverse diffusion process as a noise-adaptive posterior sampling procedure, in which the denoised channel proposal is corrected according to the pilot residual and the observation noise variance.

For the denoised estimate $\mathbf{h}_{0|t}$ obtained at timestep $t$, we introduce a weighted range-space correction~\cite{wang2022zero} as
\begin{equation}
    \hat{\mathbf{h}}_{0|t}
    =
    \mathbf{h}_{0|t}
    -
    \mathbf{\Sigma}_t \mathbf{A}^{\dagger}
    \left(\mathbf{A}\mathbf{h}_{0|t}-\mathbf{y}\right),
    \label{eq:weighted_correction}
\end{equation}
where $\mathbf{\Sigma}_t=\lambda_t \mathbf{I}$ and $0\leq \lambda_t \leq 1$. The coefficient $\lambda_t$ controls the strength of the correction imposed by the noisy pilot observation. When $\lambda_t=1$, the update reduces to the full range-space correction. When $\lambda_t<1$, the correction is relaxed to avoid injecting excessive observation noise into the reverse sampling process. Moreover, it is worth emphasizing that, in the considered OFDM pilot model, the measurement matrix \(\mathbf A\) is a sparse diagonal matrix induced by the DMRS pattern, i.e., \(\mathbf A=\operatorname{Diag}(\operatorname{vec}(\mathbf P))\). Therefore, \(\mathbf A^\dagger\) is implemented by element-wise inversion on the nonzero pilot entries and zero masking elsewhere, requiring no SVD or dense matrix inversion. Hence, the computation of $\mathbf{A}^{\dagger}$ introduces only a very small computational overhead in practice.

After obtaining $\hat{\mathbf{h}}_{0|t}$, the reverse transition from $\mathbf{h}_t$ to $\mathbf{h}_{t-1}$ is written as
\begin{multline}
    \mathbf{h}_{t-1} = 
    \frac{\sqrt{\bar{\alpha}_{t-1}}\beta_t}{1-\bar{\alpha}_t}
    \hat{\mathbf{h}}_{0|t}
    +
    \frac{\sqrt{\alpha_t}(1-\bar{\alpha}_{t-1})}{1-\bar{\alpha}_t}
    \mathbf{h}_t
    + \mathbf{\Phi}_t \mathbf{z},\\
    \mathbf{z}\sim \mathcal{N}(\mathbf{0},\mathbf{I}),
    \label{eq:posterior_sampling_noisy}
\end{multline}
where $\mathbf{\Phi}_t = \gamma_t \mathbf{I}$. The above update follows the DDIM-style~\cite{song2020denoising} reverse sampling form. For the channel estimation task, it can be interpreted as an iterative refinement step: the first term introduces the pilot-corrected channel estimate, the second term preserves the dependence on the current diffusion state, and the last term controls the remaining stochastic uncertainty. Therefore, in the noisy observation case, the key is to determine the sampling coefficient $\mathbf{\Phi}_t$ according to the observation noise level.

\begin{algorithm}[!t]
\caption{DANCE: Diffusion-based Noise-Adaptive Null-space Channel Estimation}
\label{alg:dance_alg}
\begin{algorithmic}[1]
\REQUIRE Sparse DMRS pilot pattern \(\mathbf P\), received pilot grid
\(\mathbf Y\), noise variance \(\sigma_y^2\), pre-trained denoising
network \(\epsilon_\theta\), noise schedule \(\{\beta_t\}_{t=1}^{T}\), channel label $c$, CFG weight $w$.
\ENSURE Estimated channel \(\hat{\mathbf h}\).

\STATE Vectorize the OFDM pilot model and define the nonzero DMRS support:
\[
\mathbf p=\operatorname{vec}(\mathbf P), \qquad
\mathbf y=\operatorname{vec}(\mathbf Y), \qquad
\Omega=\{i\mid p_i\neq 0\}.
\]
The measurement operator is induced by
\[
\mathbf A=\operatorname{Diag}(\mathbf p).
\]

\STATE Since \(\mathbf A\) is a sparse diagonal matrix, its
Moore--Penrose pseudo-inverse is also diagonal:
\[
[\mathbf A^\dagger]_{ij}=0,\quad i\neq j, \qquad
[\mathbf A^\dagger]_{ii}=
\begin{cases}
\dfrac{1}{p_i}, & i\in\Omega,\\[2mm]
0, & i\notin\Omega.
\end{cases}
\]

\STATE Initialize: \(\mathbf h_T\sim\mathcal N(\mathbf 0,\mathbf I)\). For each reverse step $t$, sample \(\mathbf z_t\sim\mathcal N(\mathbf 0,\mathbf I)\).
\STATE Compute: \(\alpha_t=1-\beta_t\),
\(\bar{\alpha}_t=\prod_{i=1}^{t}\alpha_i\), and
\(\sigma_t^2=\beta_t\), \(t\in[T]\).

\FOR{\(t=T\) to \(1\)}
\STATE \(\triangleright\) \textbf{Denoising and reconstruction:}
\[
\left[
\begin{aligned}
\tilde{\mathbf{\epsilon}}_\theta(\mathbf h_t,t,c)
&=
(1+w)\mathbf{\epsilon}_\theta(\mathbf h_t,t,c)
-
w\mathbf{\epsilon}_\theta(\mathbf h_t,t,\varnothing),\\[0.2em]
\mathbf h_{0|t}
&=
\frac{1}{\sqrt{\bar{\alpha}_t}}(\mathbf{h}_t - {\sqrt{1-\bar{\alpha}}_t}\tilde{\mathbf{\epsilon}}_\theta(\mathbf{h}_t, t,c)).
\end{aligned}
\right.
\]

\STATE \(\triangleright\) \textbf{DANCE coefficient calculation:}
\[
\hspace*{-1.5em}
\left[
\begin{aligned}
\mathbf{\Sigma}_t = \lambda_t \mathbf I, \quad \lambda_t =
\begin{cases}
1, &
\sigma_t\geq
\dfrac{\sqrt{\bar{\alpha}_{t-1}}\beta_t}{1-\bar{\alpha}_t}\sigma_y,
\\[-0.2em]
\dfrac{\sigma_t(1-\bar{\alpha}_t)}
{\sqrt{\bar{\alpha}_{t-1}}\beta_t\sigma_y},
& \mathrm{otherwise},
\end{cases}
\\[0.2em]
\Phi_t^{\mathrm{DANCE}}
=
\left(
{\sigma_t}
\left[
\sigma_t^2
-
\left(
\frac{\sqrt{\bar{\alpha}_{t-1}}\beta_t}
{1-\bar{\alpha}_t}
\right)^2
\lambda_t^2\sigma_y^2
\right]
\right)^{1/2}\mathbf I.
\end{aligned}
\right.
\]

    \STATE \(\triangleright\) \textbf{OFDM sparse-diagonal pilot correction:}
    \[
    \hat{\mathbf h}_{0|t}
    =
    \mathbf h_{0|t}
    -
    \mathbf{\Sigma}_t
    \left[
    \mathbf A^\dagger(\mathbf A\mathbf h_{0|t}-\mathbf y)
    \right].
    \]

\STATE \(\triangleright\) \textbf{Noise-adaptive posterior sampling:}
\[
\hspace*{-1.4em}
\mathbf h_{t-1}
=
\frac{\sqrt{\bar{\alpha}_{t-1}}\beta_t}{1-\bar{\alpha}_t}
\hat{\mathbf h}_{0|t}
+
\frac{\sqrt{\alpha_t}(1-\bar{\alpha}_{t-1})}{1-\bar{\alpha}_t}
\mathbf h_t
+
\Phi_t^{\mathrm{DANCE}}\mathbf z_t.
\]
\ENDFOR

\STATE Output: \(\hat{\mathbf h}=\mathbf h_0\).
\end{algorithmic}
\end{algorithm}

Since $\hat{\mathbf{h}}_{0|t}$ is corrected using the noisy pilot observation $\mathbf{y}$, the AWGN component in $\mathbf{y}$ is also involved in the reverse sampling process. According to~\eqref{eq:weighted_correction}, this noise component is first scaled by the correction coefficient $\lambda_t$. Then, in~\eqref{eq:posterior_sampling_noisy}, it is further scaled by the coefficient associated with $\hat{\mathbf{h}}_{0|t}$. Therefore, the variance contribution introduced by the noisy observation can be expressed as
\begin{equation}
    \left(
    \frac{\sqrt{\bar{\alpha}_{t-1}}\beta_t}
    {1-\bar{\alpha}_t}
    \right)^2
    \lambda_t^2\sigma_y^2 .
    \label{eq:observation_variance}
\end{equation}
To keep the total stochastic variance of the reverse step consistent with the diffusion schedule, the residual sampling variance assigned to the random sampling term is given by
\begin{equation}
    \gamma_t^2
    =
    \sigma_t^2
    -
    \left(
    \frac{\sqrt{\bar{\alpha}_{t-1}}\beta_t}
    {1-\bar{\alpha}_t}
    \right)^2
    \lambda_t^2\sigma_y^2,
    \label{eq:variance_constraint}
\end{equation}
where $\sigma_t^2$ denotes the nominal sampling variance in the reverse diffusion process, according to~\eqref{reverse_equation_diffusion}.



The correction coefficient $\lambda_t$ should be sufficiently large to preserve the pilot-domain correction, while ensuring that the variance in~\eqref{eq:variance_constraint} remains non-negative. Therefore, $\lambda_t$ is selected as
\begin{equation}
    \lambda_t =
    \begin{cases}
        1, 
        & \sigma_t \geq 
        \dfrac{\sqrt{\bar{\alpha}_{t-1}}\beta_t}
        {1-\bar{\alpha}_t}\sigma_y, \\[3mm]
        \dfrac{\sigma_t(1-\bar{\alpha}_t)}
        {\sqrt{\bar{\alpha}_{t-1}}\beta_t\sigma_y},
        & \mathrm{otherwise}.
    \end{cases}
    \label{eq:lambda_t}
\end{equation}
When the observation noise becomes small, i.e., $\sigma_y^2\rightarrow 0$, the correction coefficient approaches $\lambda_t=1$, and the update naturally reduces to the full range-space correction in the noiseless case.

The variance $\gamma_t^2$ in~\eqref{eq:variance_constraint} compensates for the uncertainty introduced by the noisy pilot observation. However, in OFDM channel estimation, the purpose of the reverse process is to recover an accurate channel response rather than to maintain large sampling randomness. Excessive stochastic perturbation, especially in the later reverse steps, may weaken the refined channel structure and degrade the final estimation accuracy. Therefore, we further introduce a noise-adaptive attenuation on the residual sampling variance:
\begin{equation}
\begin{aligned}
    \tilde{\gamma}_t^2
    &=
    {\sigma_t}
    \left[
        \sigma_t^2
        -
        \left(
        \frac{\sqrt{\bar{\alpha}_{t-1}}\beta_t}
        {1-\bar{\alpha}_t}
        \right)^2
        \lambda_t^2\sigma_y^2
    \right].
\end{aligned}
    \label{eq:attenuated_gamma}
\end{equation}
Since $0<\sigma_t<1$, the attenuated variance $\tilde{\gamma}_t^2$ is smaller than $\gamma_t^2$. Moreover, as the reverse process gradually approaches the clean channel domain, $\sigma_t$ decreases, and the random perturbation is further suppressed. This design helps stabilize the final reconstruction while still retaining the necessary stochasticity in the early reverse steps.

Finally, the sampling coefficient in~\eqref{eq:posterior_sampling_noisy} is defined as
\begin{equation}
    \mathbf{\Phi}_t^{\mathrm{DANCE}} = {\tilde{\gamma}_t}\mathbf{I}.
    \label{eq:phi_t}
\end{equation}
With this coefficient, each reverse step first obtains the denoised proposal 
$\mathbf{h}_{0|t}$, then applies the noise-adaptive range-space correction 
controlled by $\mathbf{\Sigma}_t$ to form $\hat{\mathbf{h}}_{0|t}$, and finally updates 
$\mathbf{h}_{t-1}$ with the attenuated stochastic term 
$\mathbf{\Phi}_t^{\mathrm{DANCE}}\mathbf{z}$ to account for residual uncertainty. 
By iterating this procedure, DANCE reconstructs the complete channel response by 
jointly exploiting the noisy pilot observations and the learned channel prior while 
suppressing unnecessary noise injection during the reverse process.

\subsection{Network Architecture Design}

\begin{figure*}[t]
\vspace{-4pt}
\centering
\includegraphics[width=1.0\textwidth]{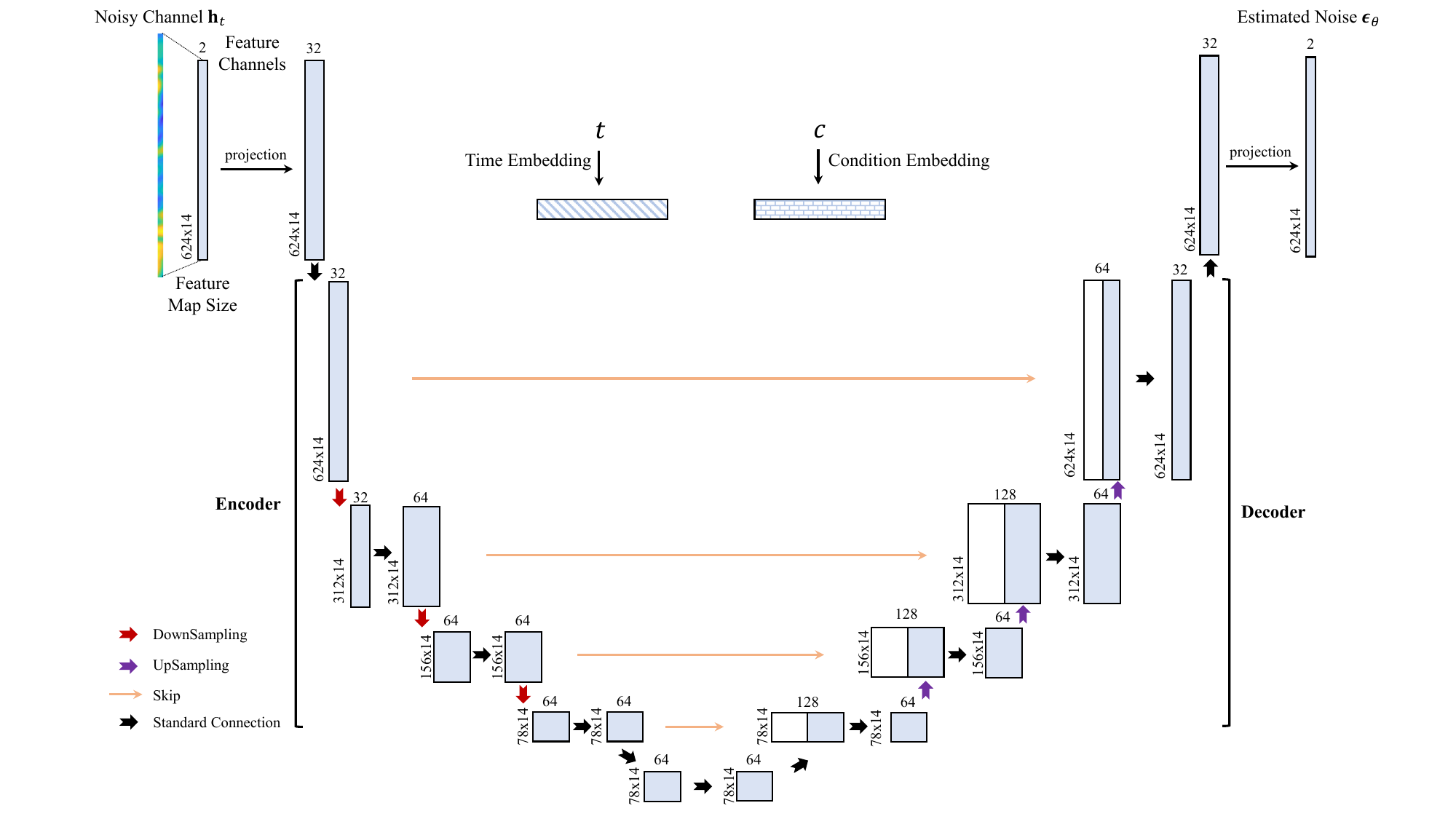}
\caption{Detailed architecture of the proposed U-Net denoising network. 
The input noisy channel sample $\mathbf{h}_t$ and the output estimated noise $\mathbf{\epsilon}_{\theta}$ are represented by two feature channels corresponding to the real and imaginary parts. 
The numbers above the blocks indicate feature channels, and the vertical labels indicate feature map sizes.
Downsampling and upsampling are performed only along the subcarrier dimension, while skip connections are used to fuse encoder and decoder features.}
\vspace{-4pt}
\label{fig:network_arch}
\end{figure*}

The denoising backbone is designed as an OFDM-tailored conditional U-Net~\cite{ronneberger2015u}, as illustrated in Fig.~\ref{fig:network_arch}. In the proposed diffusion-based channel estimator, the role of the neural network is to predict the noise component $\mathbf{\epsilon}_\theta(\mathbf{h}_t,t)$ from the noisy channel sample $\mathbf{h}_t$ at diffusion step $t$. This predicted noise is then used in the reverse denoising process to progressively recover the clean channel response.

Since the channel response in OFDM systems is complex-valued, the input channel matrix $\mathbf{h}_t$ is represented by separating its real and imaginary parts into two feature channels. Therefore, the input tensor has two input channels, corresponding to $\Re\{\mathbf{h}_t\}$ and $\Im\{\mathbf{h}_t\}$, respectively. It is worth noting that the word ``channel'' in the network architecture refers to the number of feature maps in the neural network, rather than the wireless propagation channel. The output of the network has the same two-channel format and represents the estimated noise in the real and imaginary components.

The time-step index $t$ is introduced to indicate the current noise level in the diffusion process. Specifically, different values of $t$ correspond to different corruption levels of the channel sample, and the denoising network should apply different restoration strengths at different steps. To provide this information to the network, $t$ is first mapped into a time embedding through MLP layers, and then incorporated into the intermediate feature representations. In addition, we introduce condition index $c$ to the neural network $\mathbf{\epsilon}_\theta(\mathbf{h}_t,t,c)$, which is used to represent the channel scenario or channel type. It is also processed through an embedding module and injected into the network, enabling the denoiser to adapt its prediction to different channel distributions.

The base number of feature channels in the U-Net is set to 32, and the channel multipliers are configured as $[1,2,2,2]$. Thus, the number of feature channels increases progressively in the encoder, allowing the network to extract more abstract representations as the feature resolution decreases. During downsampling, we exploit the structural property of the OFDM resource grid. Since the dimension along the subcarrier domain is much larger than that along the OFDM symbol domain, downsampling is only performed along the subcarrier dimension, while the symbol dimension remains unchanged. This design reduces computational complexity while preserving the temporal structure across OFDM symbols.

The decoder contains three upsampling stages, which progressively restore the feature maps to the original time--frequency resolution. At each corresponding resolution, skip connections are used to concatenate encoder features with decoder features. These skip connections allow fine-grained local information from shallow layers to be directly reused during reconstruction, which is important for preserving detailed time-frequency channel structures. Through this encoder-decoder design, the network can jointly capture local channel variations and global structural correlations over the OFDM resource grid.

In this work, we further adopt the classifier-free guidance mechanism~\cite{ho2022classifier} to enhance the conditional generation capability of the diffusion denoiser. The purpose of this mechanism is to allow a single network to learn both conditional and unconditional denoising behaviors. During training, the condition index $c$ is randomly dropped with probability $p_{\mathrm{uncond}}$ and replaced by a null token $\varnothing$. In this way, the same denoising network learns two types of predictions: the conditional prediction $\mathbf{\epsilon}_\theta(\mathbf{h}_t,t,c)$, where $c$ denotes the channel-scenario condition such as the TDL/CDL profile, and the unconditional prediction $\mathbf{\epsilon}_\theta(\mathbf{h}_t,t,\varnothing)$, where $\varnothing$ denotes the null condition without using the channel label.

During inference, the denoising network is evaluated twice at each reverse step, once with the condition label $c$ and once with the null condition $\varnothing$. The two predicted noise terms are then linearly combined as
\begin{equation}
\tilde{\mathbf{\epsilon}}_\theta(\mathbf{h}_t,t,c)
=
(1+w)\mathbf{\epsilon}_\theta(\mathbf{h}_t,t,c)
-
w\mathbf{\epsilon}_\theta(\mathbf{h}_t,t,\varnothing),
\label{eq:cfg}
\end{equation}
where $w\geq 0$ is the guidance weight. When $w=0$, the guided prediction reduces to the standard conditional prediction. When $w$ increases, the difference between the conditional and unconditional predictions is amplified, which strengthens the influence of the channel condition on the reverse sampling trajectory.

For the OFDM channel estimation task, this guidance mechanism can be understood as a way of steering the denoising process toward the channel distribution specified by the condition label. As a result, the network can better distinguish different channel scenarios while still sharing the same set of model parameters. This improves the adaptability of the proposed estimator under different channel types and helps the reverse diffusion process generate channel estimates that are more consistent with the target propagation environment.

The complete procedure of the proposed diffusion-based OFDM channel estimation DANCE, including the range-space correction and noise-adaptive posterior sampling, is summarized in Algorithm~\ref{alg:dance_alg}.

\section{Experimental Results}
This section presents numerical experimental results to evaluate the performance of DANCE, the proposed diffusion model-based channel estimation method. We first describe the experimental environment and configuration, followed by a comprehensive performance analysis of the proposed estimator across multiple channel scenarios, including different SNRs, DMRS configurations, Doppler frequency shifts, and train--test distribution mismatches.

\subsection{Experimental Setup}
For the experiments, the training and testing channel datasets are generated using the 5G NR Tapped Delay Line (TDL) (via the \texttt{nrTDLChannel} function) and Clustered Delay Line (CDL) (via the \texttt{nrCDLChannel} function) channel models in MATLAB~\cite{3gpp_tr_38.901_2022}. The downlink transmission employs an OFDM frame structure with the following parameters: an FFT size of 1024, a sampling rate of 15.36 MHz, a subcarrier spacing of 15 kHz, a cyclic prefix length of 144, and 14 OFDM symbols per slot. A total of 52 resource blocks are allocated, corresponding to $N_{RE}=624$ resource elements. Both the base station and user equipment are configured with a single antenna. All pilot symbols are randomly selected from a Quadrature Phase Shift Keying (QPSK) set of symbols and are normalized to unit power. In the experimental setup, multiple channel datasets with distinct characteristics are generated. To enhance the model's generalization capability and inference efficiency, a conditional diffusion model is adopted for both training and inference. Specifically, we model the channel type (e.g., CDL-A, TDL-C) as a discrete condition and train a diffusion denoiser that captures the channel distribution conditioned on this label. Each dataset consists of 50,000 samples for training and 100 independent samples for testing.\textbf{}

During training, the number of epochs is set to 128 with a batch size of 128. The Adam optimizer is employed with a learning rate of $2 \times 10^{-4}$. All experiments are conducted on a single NVIDIA L40 GPU with 48 GB memory. For the diffusion model configuration, a linear noise schedule is adopted with a total of $T=1000$ diffusion steps during training. During inference, unless otherwise specified, the reverse sampling process is performed with 200 sampling steps to balance estimation accuracy and computational complexity. For classifier-free guidance, the guidance weight \(w\) is fixed at \(4.0\) throughout all experiments. For evaluation, the Normalized Mean Squared Error (NMSE) is used as the performance metric:
\begin{equation}
\mathrm{NMSE}=\mathbb{E}\!\left[\frac{\left\|\hat{\mathbf{h}}-\mathbf{h}\right\|_2^{2}}{\left\|\mathbf{h}\right\|_2^{2}}\right],
\end{equation}
where $\mathbf{h}$ denotes the ground-truth channel vector and $\hat{\mathbf{h}}$ denotes the estimated channel vector.


For benchmarking, we compare our proposed method, DANCE, against the following four schemes:

\begin{enumerate}
    \item \textbf{MMSE Estimator:} The structure of the MMSE estimator is illustrated in Fig.~\ref{fig:MMSE}. The process begins with an LS estimate of the observed signal. Subsequently, frequency-domain filtering is applied only at the positions of the DMRS pilots. Following this, since DMRS pilots are present only on specific OFDM symbols, time-domain interpolation (TDI) is performed. Finally, to account for DMRS pilots occupying only certain subcarrier positions, frequency-domain interpolation (FDI) is carried out. The Wiener coefficients used for both the filtering and interpolation stages are computed in a similar manner. The requisite correlation information is obtained through prior statistical analysis of the training dataset.
    \begin{figure}[H]
    \vspace{-4pt} 
    \centering
    \includegraphics[width=\columnwidth]{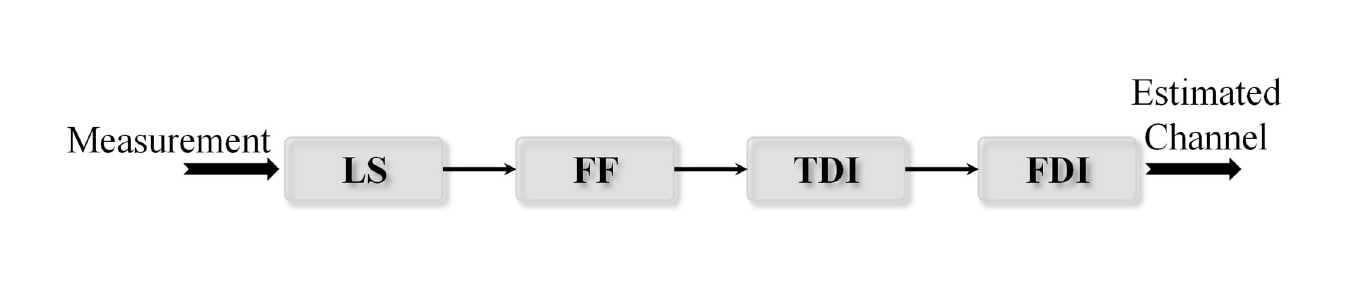}
    \caption{Block diagram of MMSE.}
    \vspace{-4pt} 
    \label{fig:MMSE}
    \end{figure}

    \item \textbf{MATLAB 5G Toolbox Channel Estimator:} The \texttt{nrChannelEstimator}~\cite{matlab_nrchannelestimate} function from the MATLAB 5G Toolbox is employed. This estimator performs pilot averaging and interpolation, supports TDL and CDL channel models, and integrates functionalities for noise estimation and delay-spread approximation.

    \item \textbf{DPS:} The Diffusion Posterior Sampling (DPS)~\cite{chung2022diffusion} method adopts gradient calculation of the distance between observation and prediction to guide the reverse diffusion trajectory. It requires backpropagation through the neural network while maintaining accuracy and versatility in various scenarios, resulting in high computational cost and slow inference.

    \item \textbf{DMPS:} The Diffusion Model Based Posterior Sampling (DMPS)~\cite{meng2022diffusion} method creates an analytical closed-form approximation for the likelihood score, enabling significantly faster inference than DPS in linear inverse problems at the expense of estimation accuracy.
\end{enumerate}

\begin{figure*}[!t]
\centering
\setlength{\tabcolsep}{0.5mm} 

\begin{tabular}{@{}cccc@{}}
\panel{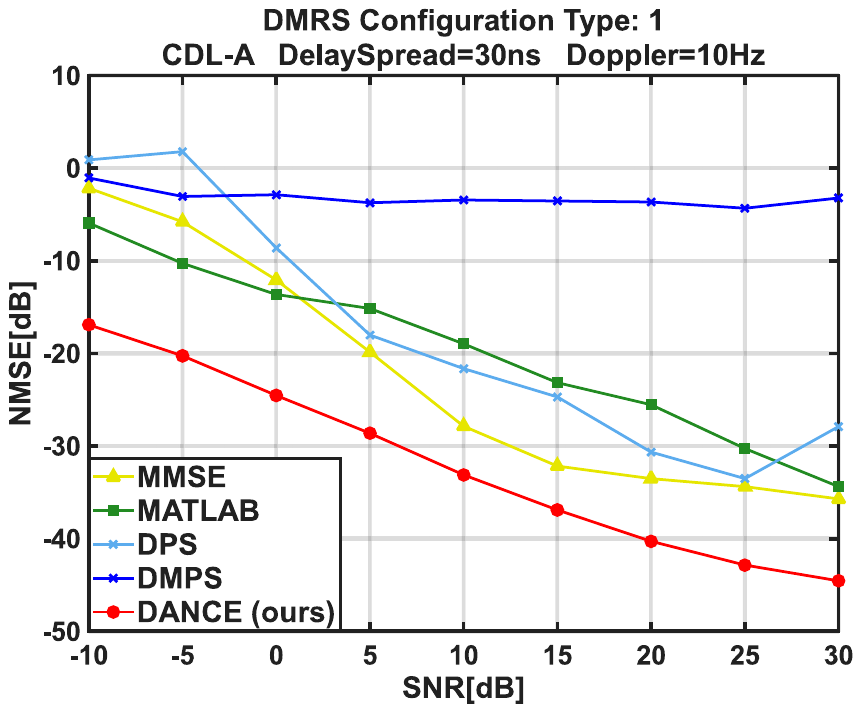} &
\panel{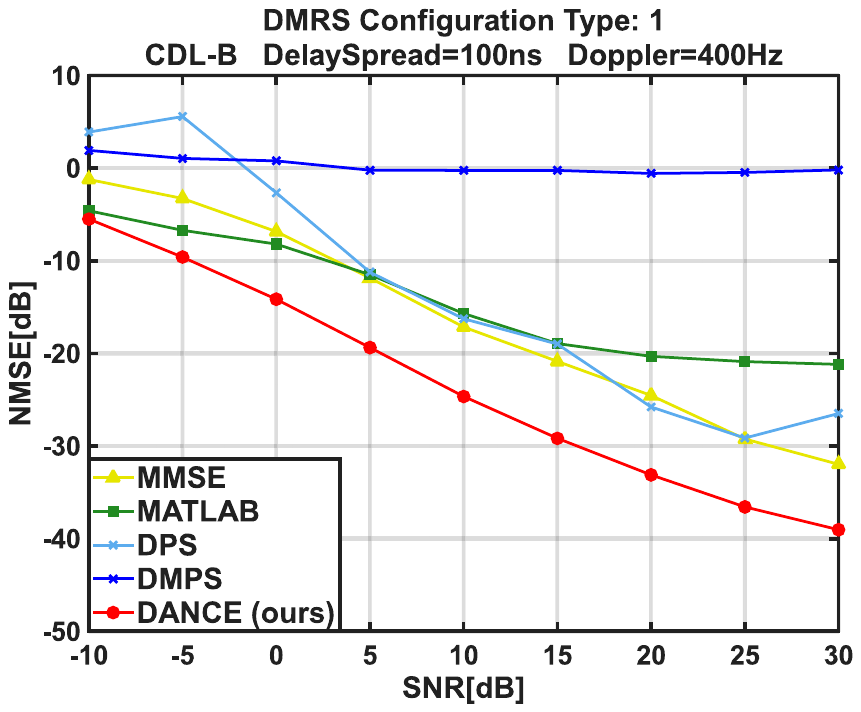} &
\panel{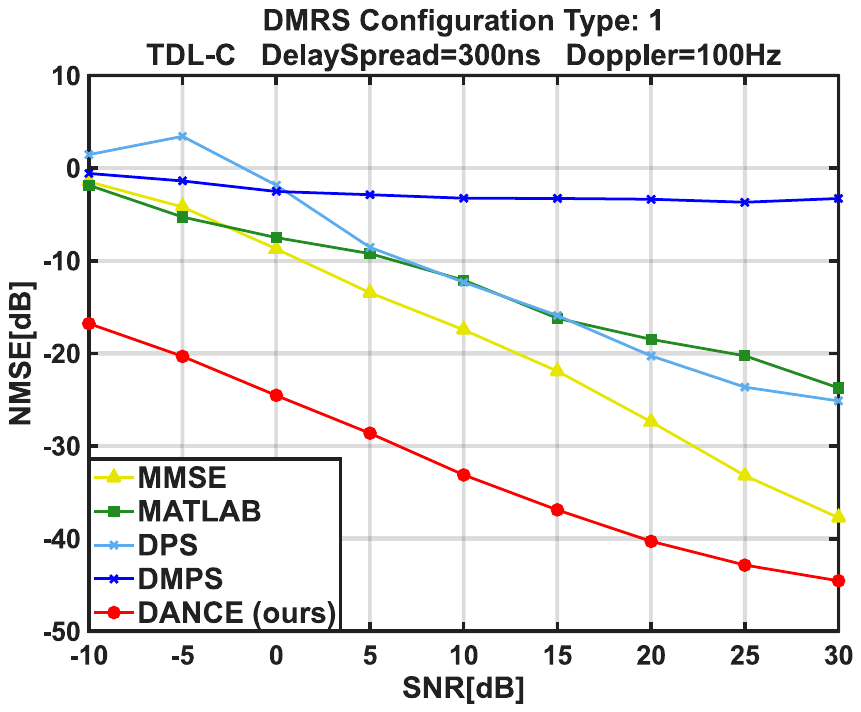} &
\panel{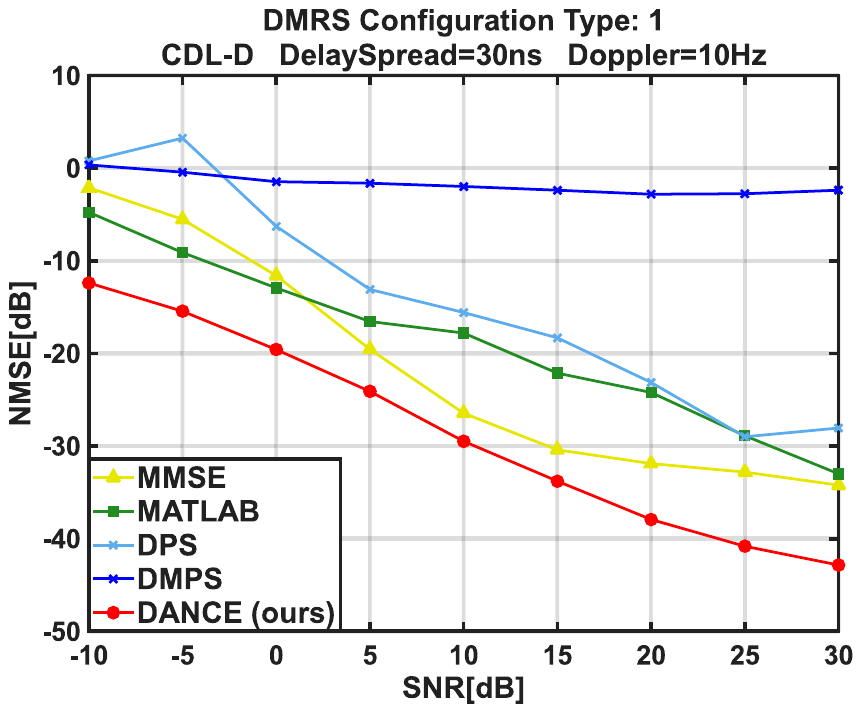} \\[0mm]
\panel{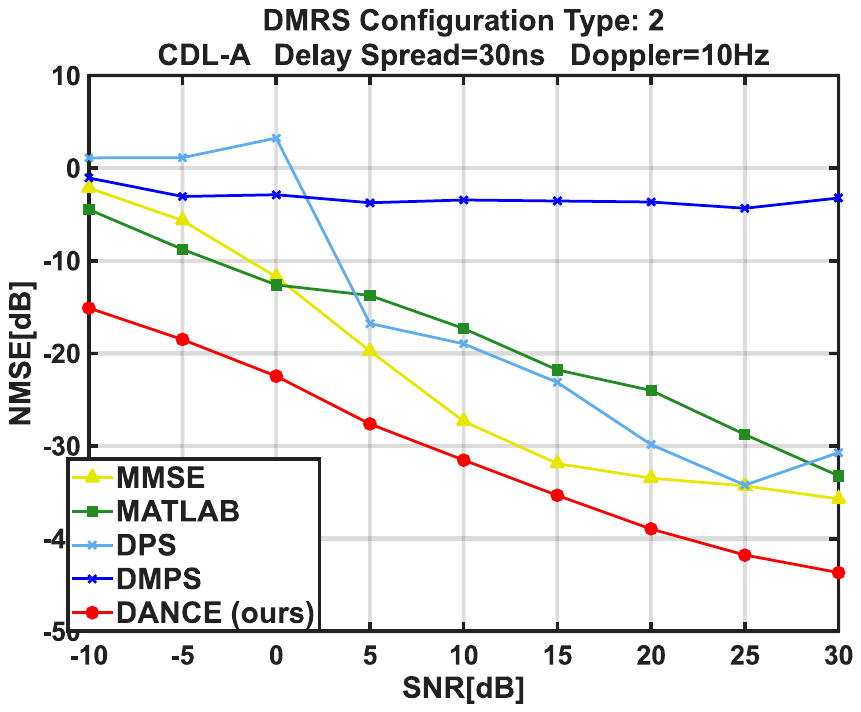} &
\panel{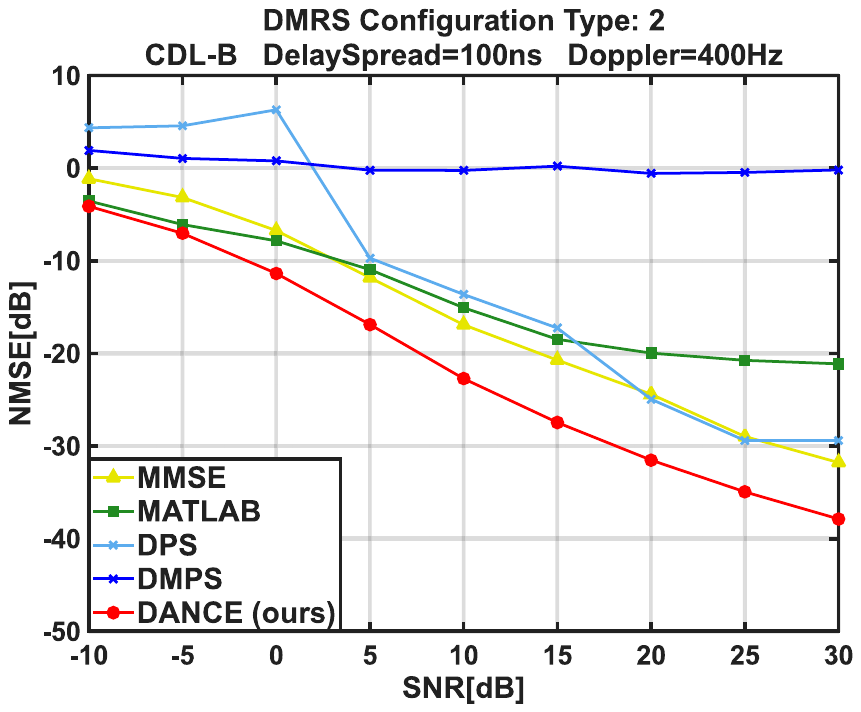} &
\panel{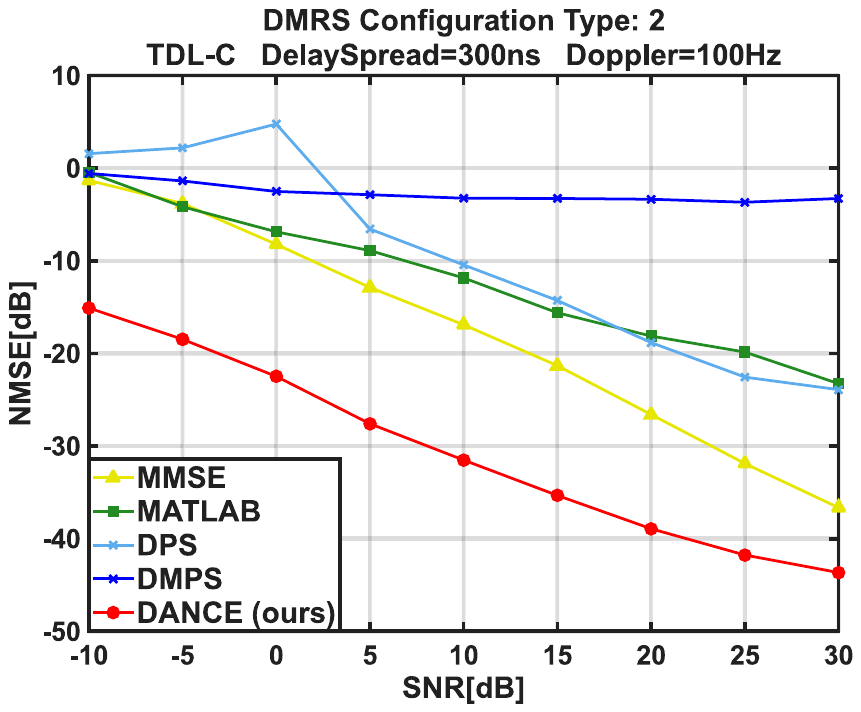} &
\panel{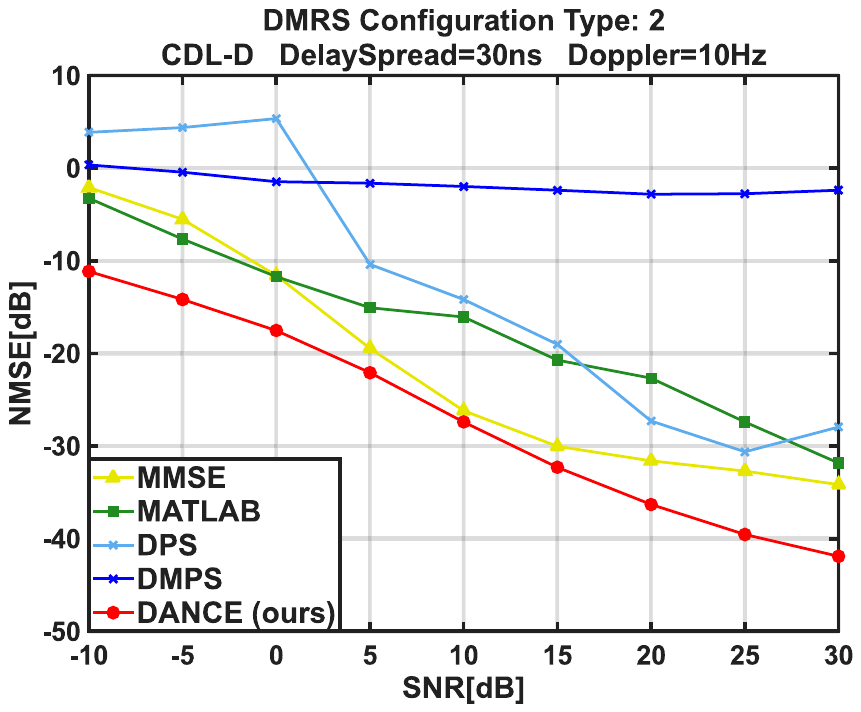} \\
\end{tabular}

\vspace{-0.5mm}
\caption{NMSE versus SNR performance under different channel environments and DMRS configuration types.}
\label{fig:NMSE}
\end{figure*}

\subsection{Standard Experiments}

We begin by specifying the channel environments and pilot configurations under test. Four distinct channel datasets are randomly generated, labeled \texttt{CDLA30-10}, \texttt{CDLB100-400}, \texttt{CDLD30-10}, and \texttt{TDLC300-100}. In each label, the first numerical value denotes the root mean square (RMS) delay spread in nanoseconds, while the second value represents the maximum Doppler shift in Hertz. For example, \texttt{TDLC300-100} corresponds to a TDL-C channel with an RMS delay spread of 300 ns and a maximum Doppler shift of 100 Hz. The configuration of three DMRS symbols per slot was maintained throughout. Additionally, we assessed methods' performance under two Demodulation Reference Signal Configuration Types (CTs)~\cite{channels2020nr}. The relevant configuration parameters are summarized in Table~\ref{tab:Config_std_exp}.
\begin{table}[H]
\caption{Configurations for standard experiments.}
\centering
\setlength{\tabcolsep}{8pt}
\renewcommand{\arraystretch}{1.15}

\begin{tabular}{|>{\columncolor{gray!20}}c|c|c|c|c|}
\hline
\rowcolor{gray!35}
\textbf{Profile} & \textbf{RMS (ns)} & \textbf{Doppler (Hz)} & \textbf{CT} & \textbf{DMRS symbols}\\
\hline
\textbf{CDL-A} & 30  & 10  & 1/2 &3\\ \hline
\textbf{CDL-B} & 100 & 400 & 1/2 &3\\ \hline
\textbf{TDL-C} & 300 & 100 & 1/2 &3\\ \hline
\textbf{CDL-D} & 30  & 10  & 1/2 &3\\ \hline
\end{tabular}

\label{tab:Config_std_exp}
\end{table}

By experimental comparison as illustrated in Fig.~\ref{fig:NMSE}, it is evident that under all four channel environments and both DMRS configuration types, our method DANCE  outperforms the baseline approaches. The DPS method exhibits high and unstable NMSE values, mainly due to the difficulty of accurately guiding the reverse process through gradient-based posterior updates. The DMPS method also suffers from severe performance degradation under sparse pilot configurations, with its NMSE tending to saturate around 0 dB. This indicates that DMPS cannot provide reliable channel reconstruction when the pilot density is low. However, as the pilot density increases, more observation constraints become available, and the performance of DMPS can be improved accordingly. Nevertheless, it still yields higher NMSE than DANCE in the considered low-pilot-density OFDM scenarios. The MMSE estimator and the MATLAB 5G Toolbox estimator show similar decreasing trends as the SNR increases, but their performance is still limited by the sparse pilot observations and the accuracy of statistical assumptions. Overall, these results indicate that DANCE achieves consistently lower NMSE than the compared baselines across the considered channel conditions and DMRS configurations, suggesting improved robustness to channel and pilot-pattern variations.


Meanwhile, since the previous experiments maintained the configuration of three DMRS symbols per slot, we subsequently tested the impact of the number of DMRS symbols per slot on the proposed method under fixed channel conditions. The specific corresponding configurations are listed in Table~\ref{tab:DMRS symbols}.

\begin{table}[H]
\caption{Configurations for different numbers of DMRS symbols.}
\centering
\setlength{\tabcolsep}{8pt}
\renewcommand{\arraystretch}{1.15}

\begin{tabular}{|>{\columncolor{gray!20}}c|c|c|c|c|}
\hline
\rowcolor{gray!35}
\textbf{Profile} & \textbf{RMS (ns)} & \textbf{Doppler (Hz)} & \textbf{CT} & \textbf{DMRS symbols} \\
\hline
\textbf{CDL-A} & 30  & 10  & 1 & 2/3/4 \\ \hline
\end{tabular}

\label{tab:DMRS symbols}
\end{table}

\begin{figure}[H]
\centering
\includegraphics[width=1.0\columnwidth, trim=0mm 70mm 0mm 70mm, clip]{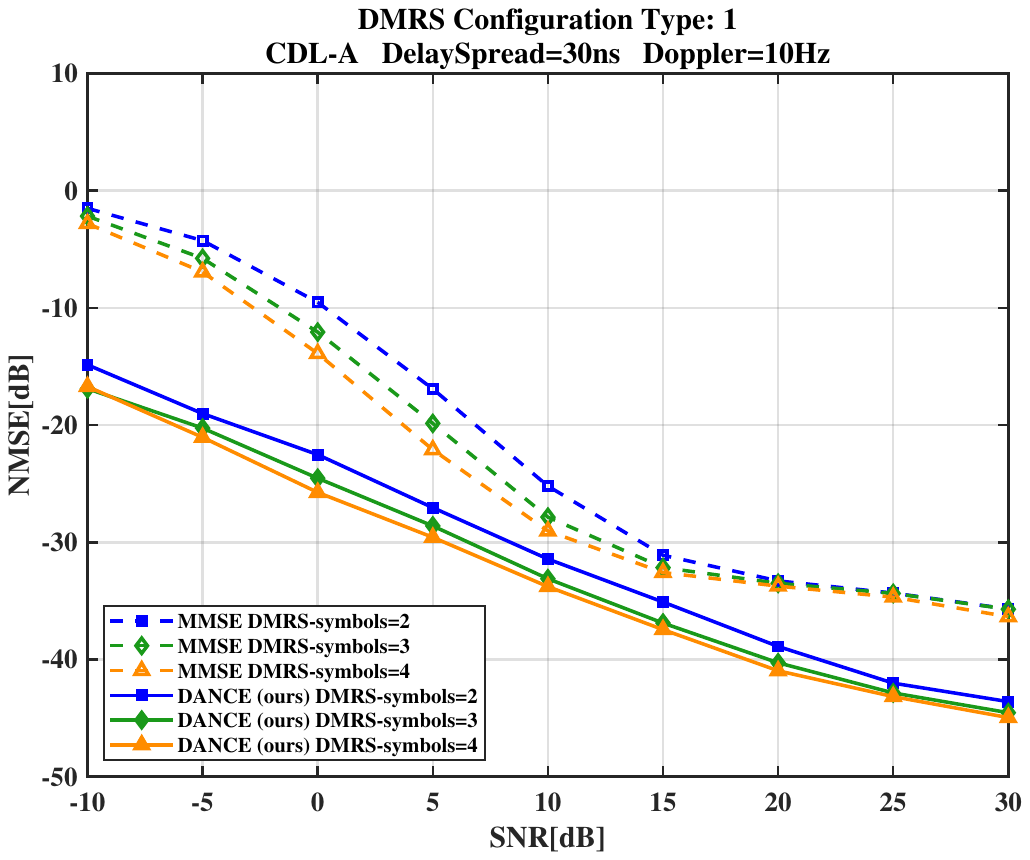}
\caption{NMSE versus SNR performance under different numbers of DMRS symbols.}
\label{fig:DMRS_symbol}
\end{figure}

\begin{figure*}[!t]
\centering
\setlength{\tabcolsep}{0.5mm} 

\begin{tabular}{@{}cccc@{}}
\panel{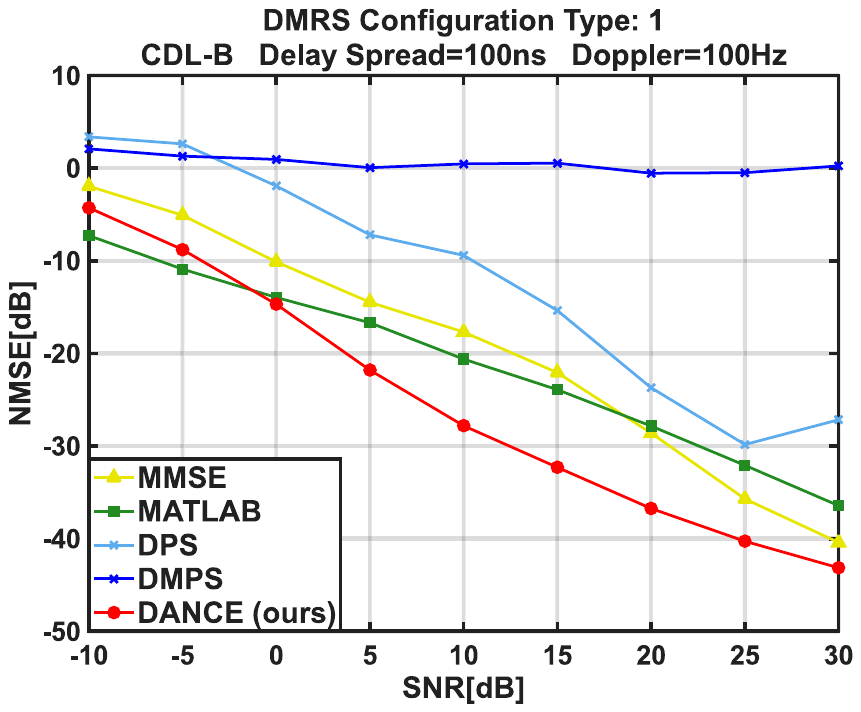} &
\panel{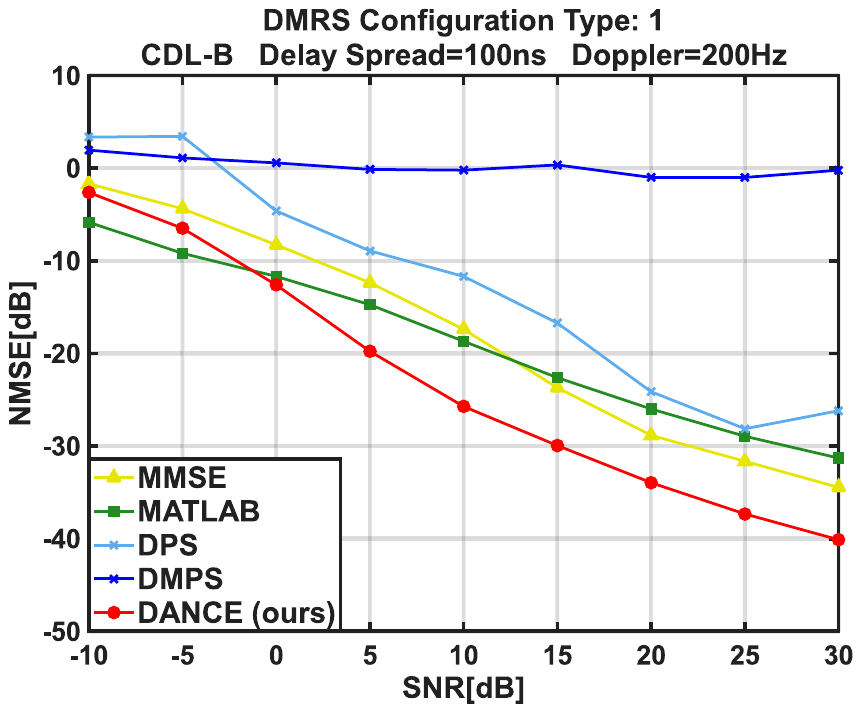} &
\panel{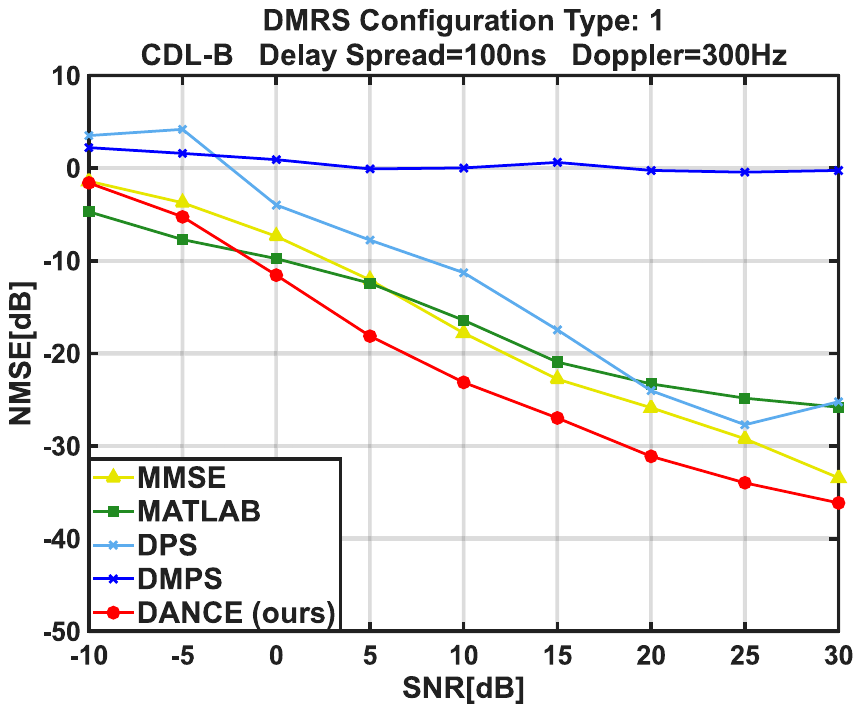} &
\panel{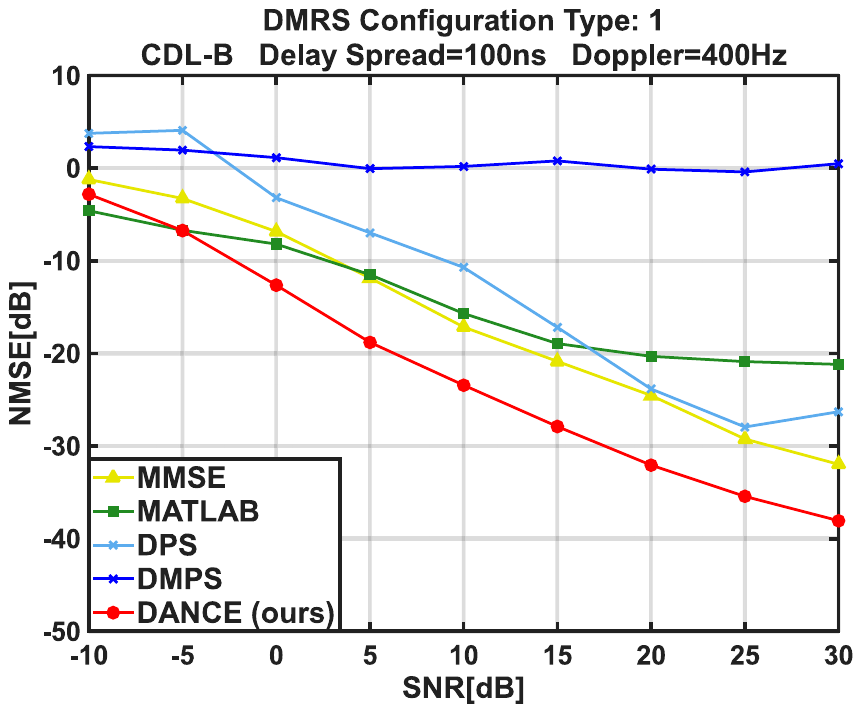}
\end{tabular}

\vspace{-0.5mm}
\caption{NMSE versus SNR performance under different Doppler frequency shifts.}
\label{fig:NMSE_doppler_shift}
\end{figure*}

The experimental results on different numbers of DMRS symbols are shown in Fig.~\ref{fig:DMRS_symbol}. As the number of DMRS symbols increases, the performance of both DANCE and the MMSE estimator improves steadily. This improvement stems from the fact that a larger number of symbols corresponds to a higher pilot density, which provides the receiver with more measurement information and thus simplifies channel reconstruction. The results further show that DANCE maintains lower NMSE than the MMSE estimator across different numbers of DMRS symbols, indicating robustness to changes in pilot density.


Furthermore, for high-mobility scenarios, estimation accuracy remains a critical metric. A larger Doppler frequency shift generally corresponds to a higher mobility level. On this basis, we evaluate the performance of DANCE under different Doppler frequency shifts. The specific channel testing scenarios are listed in Table~\ref{tab:doppler shift}.

\begin{table}[H]
\caption{Configurations for different Doppler frequency shifts.}
\centering
\setlength{\tabcolsep}{7pt}
\renewcommand{\arraystretch}{1.15}

\begin{tabular}{|>{\columncolor{gray!20}}c|c|c|c|c|}
\hline
\rowcolor{gray!35}
\textbf{Profile} & \textbf{RMS (ns)} & \textbf{Doppler (Hz)} & \textbf{CT} & \textbf{DMRS symbols} \\
\hline
\textbf{CDL-B} & 100  & 100/200/300/400  & 1 & 3 \\ \hline
\end{tabular}

\label{tab:doppler shift}
\end{table}

The corresponding experimental results are shown in Fig.~\ref{fig:NMSE_doppler_shift}. We evaluate the performance of DANCE against the baselines under the CDL-B channel configurations \texttt{CDLB100-100}, \texttt{CDLB100-200}, \texttt{CDLB100-300}, and \texttt{CDLB100-400}. 
The results indicate that DANCE maintains lower NMSE than the compared baselines across the considered Doppler frequency shifts, suggesting improved robustness under mobility-induced channel variations.


\subsection{Generalization Experiments}
In this section, we primarily examine the generalization capability of the diffusion model-based channel estimation method DANCE under scenarios where the distributions of the training data and the testing data are significantly different.

We consider two types of generalization, namely \emph{In-Distribution (In-D) Generalization} and \emph{Out-of-Distribution (OoD) Generalization}~\cite{pratik2025requestnet}. In-D generalization refers to scenarios where the distributions of the testing data and the training data are not identical but remain structurally similar. For example, the model is trained on the TDL-C channel type but tested on the TDL-D or TDL-E channel types. In contrast, OoD generalization refers to scenarios where the testing data exhibit more substantial structural differences from the training distribution, such as training on a TDL-C  channel model and testing on CDL-A or CDL-B channel models.

\begin{table}[H]
\caption{Configurations for generalization experiments.}
\centering
\setlength{\tabcolsep}{14pt}
\renewcommand{\arraystretch}{1.15}

\begin{tabular}{|>{\columncolor{gray!20}}c|c|c|c|}
\hline
\rowcolor{gray!35}
\textbf{Category} & \textbf{Profile} & \textbf{RMS (ns)} & \textbf{Doppler (Hz)}\\
\hline
\textbf{In-D} & TDL-D  & 30  & 100\\ \hline
\textbf{In-D} & TDL-E  & 100 & 100\\ \hline
\textbf{OoD}  & CDL-A  & 30  & 100\\ \hline
\textbf{OoD}  & CDL-B  & 100  & 100\\ \hline
\end{tabular}

\label{tab:Config_gen_exp}
\end{table}

In this subsection, the specific experimental configurations designed to evaluate the robustness of the proposed method under train--test distribution mismatch are detailed in Table~\ref{tab:Config_gen_exp}. Specifically, our benchmark training dataset is exclusively \texttt{TDLC300-100}. That is, for both in-distribution and out-of-distribution generalization tests, the model is trained only on the \texttt{TDLC300-100} dataset. For in-distribution generalization evaluation, the test datasets are \texttt{TDLD30-100} and \texttt{TDLE100-100}. For out-of-distribution generalization evaluation, the test datasets are \texttt{CDLA30-100} and \texttt{CDLB100-100}. In this series of experiments, since MATLAB \texttt{nrChannelEstimator} does not rely on learned or training-derived channel priors, it is not affected by train--test distribution mismatch in the same sense as learning-based methods. We therefore report it as a reference conventional estimator on each target test channel. In contrast, the MMSE estimator is capable of controlling the consistency between prior statistical information and the test dataset, thereby enabling such experiments to be conducted.
All diffusion-based methods evaluated, including DPS, DMPS, and the proposed DANCE, are tested under the same generalization conditions, where the training and test datasets are intentionally different. This allows a fair comparison of their robustness and adaptability to \emph{In-D} and \emph{OoD} scenarios.

\begin{figure*}[!t]
\centering
\setlength{\tabcolsep}{0.5mm} 

\begin{tabular}{@{}cccc@{}}
\panel{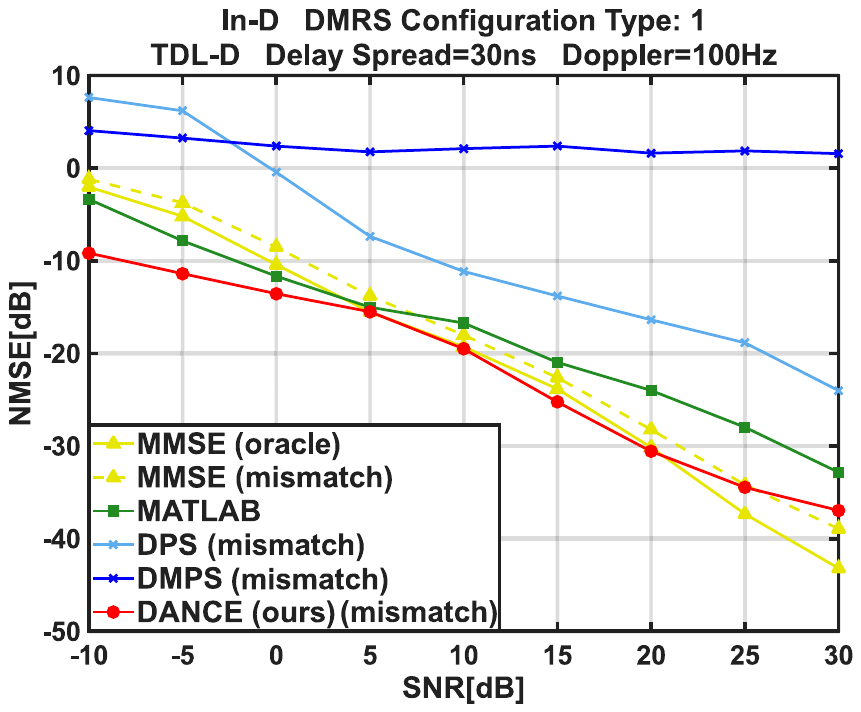} &
\panel{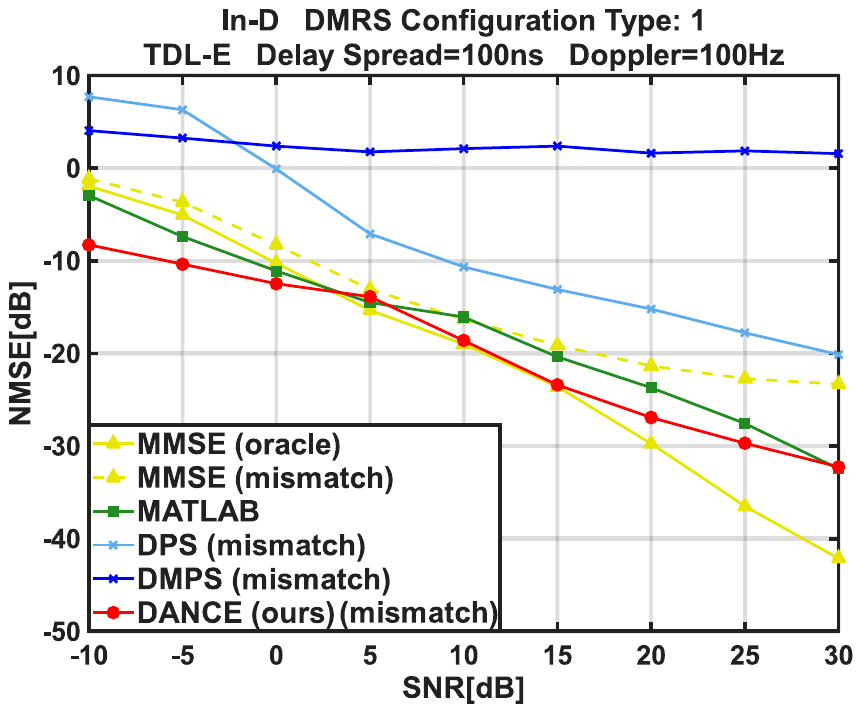} &
\panel{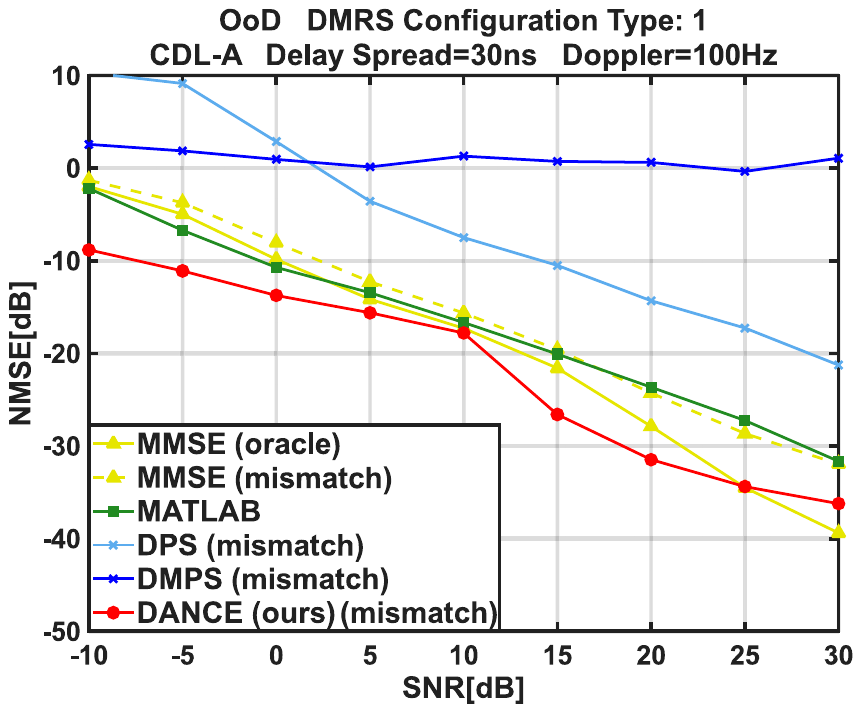} &
\panel{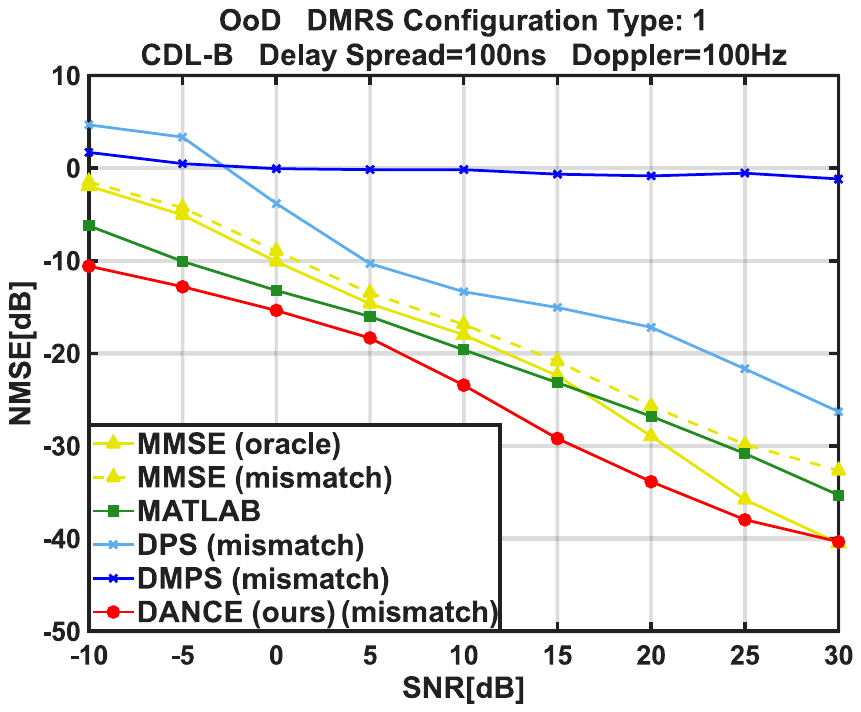}
\end{tabular}

\vspace{-0.5mm}
\caption{NMSE versus SNR performance under generalization scenarios.}
\label{fig:NMSE_generalization}
\end{figure*}

The experimental results for this part are presented in Fig.~\ref{fig:NMSE_generalization}. As can be observed, DANCE achieves performance comparable to that of the MMSE estimator with perfect prior knowledge, even when the training and test dataset distributions are not identical. Particularly under low SNR conditions, the proposed method significantly outperforms MMSE. Although it underperforms MMSE in certain scenarios, this outcome is aligned with expectations. The MMSE baseline benefits from statistical prior information matched to the target test dataset, whereas DANCE uses a fixed learned prior without explicit alignment or fine-tuning to each testing scenario. Therefore, the observed gap in some cases is reasonable under the considered distribution-mismatch setting. The DPS method follows the same trend as our method while maintaining a higher NMSE level. The DMPS method exhibits a saturated NMSE level around 0 dB, which suggests that DMPS requires a higher pilot density for higher estimation and generalization accuracy. These results indicate that DANCE is more robust to train--test distribution mismatch than the diffusion posterior-sampling baselines in the considered scenarios, while remaining competitive with the oracle MMSE reference that uses target-matched statistical information. Compared with the MATLAB 5G Toolbox estimator and the mismatched MMSE estimator, DANCE also achieves lower NMSE in most tested settings. This suggests that the learned diffusion prior, combined with the proposed noise-adaptive correction, can provide useful generalization capability across different channel models without target-specific fine-tuning.



\subsection{Ablation Study}

In this subsection, we conduct an ablation study to evaluate the effectiveness of the proposed noise-adaptive posterior correction strategy. Specifically, we compare the baseline diffusion estimator without the proposed correction and the enhanced version with the correction under three representative channel scenarios, namely \texttt{TDLA30-10}, \texttt{TDLB100-400}, and \texttt{TDLC300-100}. All experiments are carried out under DMRS configuration type 1 and 3 DMRS symbols per slot. The specific channel testing scenarios are listed in Table~\ref{tab:ablation_study_correction}.

\begin{table}[H]
\caption{Configurations for the ablation study on noise-adaptive posterior correction.}
\centering
\setlength{\tabcolsep}{10pt}
\renewcommand{\arraystretch}{1.15}

\begin{tabular}{|>{\columncolor{gray!20}}c|c|c|c|}
\hline
\rowcolor{gray!35}
\textbf{Profile} & \textbf{RMS (ns)} & \textbf{Doppler (Hz)} & \textbf{Correction} \\
\hline
\textbf{TDLA30-10}   & 30  & 10   & \ding{52}/\ding{56}\\ \hline
\textbf{TDLB100-400} & 100 & 400  & \ding{52}/\ding{56}\\ \hline
\textbf{TDLC300-100} & 300 & 100  & \ding{52}/\ding{56}\\ \hline
\end{tabular}

\label{tab:ablation_study_correction}
\end{table}



\begin{figure}[t]
\centering
\includegraphics[
    width=0.85\columnwidth,
    trim={0mm 60mm 0mm 60mm},
    clip
]{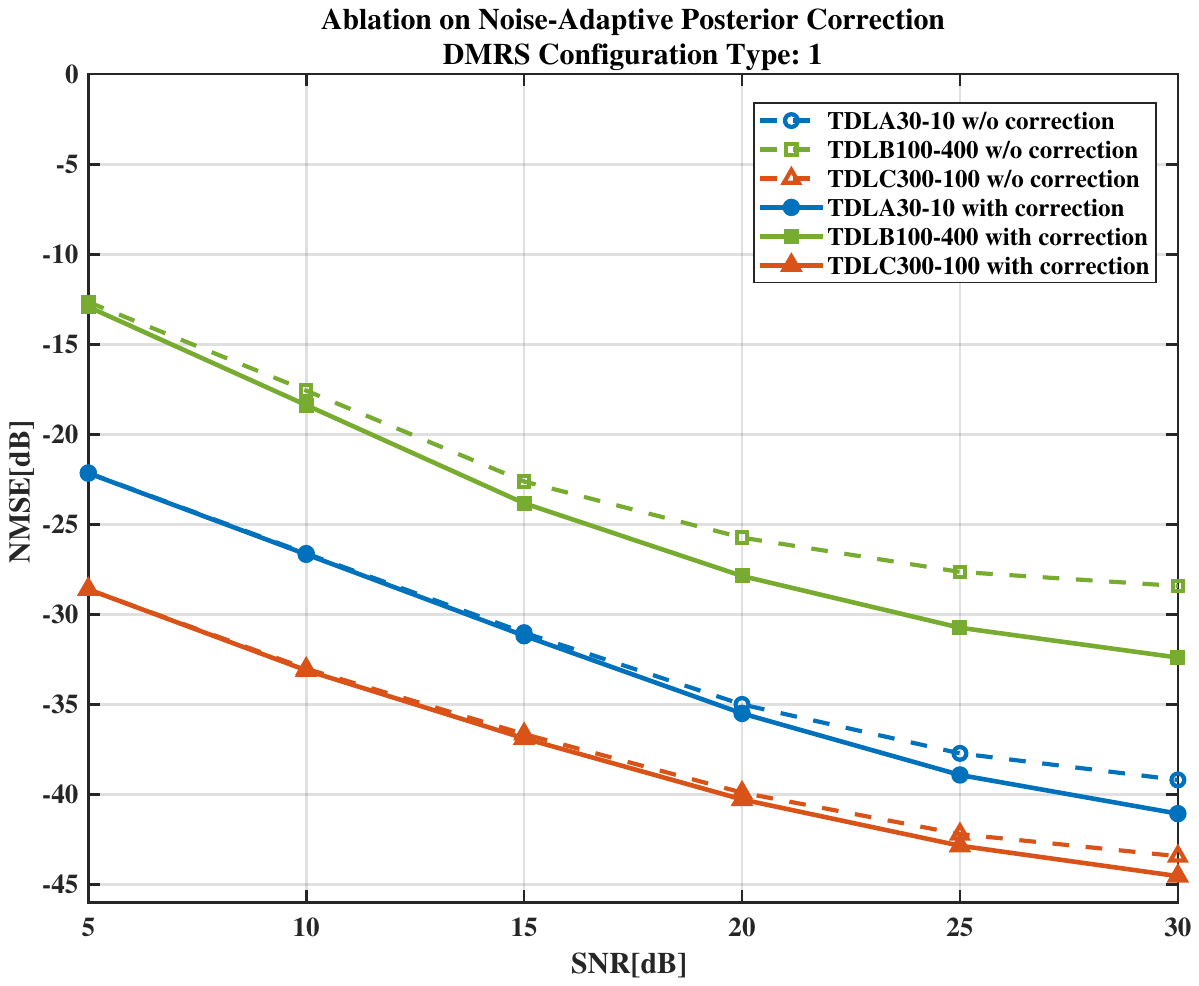}
\caption{Ablation study on the proposed noise-adaptive posterior correction under DMRS configuration type 1 and 3 DMRS symbols per slot. Dashed curves correspond to the baseline diffusion estimator without the proposed correction, while solid curves correspond to the estimator with the proposed correction. Different colors denote different channel scenarios.}
\vspace{-6pt}
\label{fig:ablation_correction}
\end{figure}

The corresponding results are shown in Fig.~\ref{fig:ablation_correction}. It can be observed that the proposed correction consistently improves the NMSE performance across the three channel scenarios, while the improvement is particularly evident in the high-SNR regime. In the low-SNR region, the performance gap between the two versions is relatively limited, since the estimation accuracy is mainly constrained by the severe observation noise. In contrast, as the SNR increases, the advantage of the proposed correction becomes increasingly pronounced.

More specifically, for the \texttt{TDLB100-400} channel, the proposed correction yields clear gains of approximately $2.13$ dB, $3.10$ dB, and $3.98$ dB at SNRs of $20$, $25$, and $30$ dB, respectively. For the \texttt{TDLA30-10} channel, the corresponding gains are about $0.50$ dB, $1.20$ dB, and $1.87$ dB, while for the \texttt{TDLC300-100} channel, the gains reach about $0.38$ dB, $0.65$ dB, and $1.11$ dB, respectively. These results clearly demonstrate that the proposed correction is beneficial for improving the final reconstruction quality, especially when the pilot observations are relatively reliable.

The above phenomenon is consistent with the design motivation of the proposed method. In the reverse sampling process, the noise-adaptive posterior correction jointly adjusts the observation-consistency correction strength and the residual sampling variance according to the observation noise level. When the SNR is high, the pilot observations become more reliable, and the proposed strategy can more effectively exploit such information while suppressing unnecessary stochastic perturbations in the late reverse steps. As a result, the reconstructed channel becomes more accurate, leading to a more noticeable NMSE improvement. Overall, this ablation study verifies that the proposed noise-adaptive posterior correction is an effective component of the proposed diffusion-based channel estimation framework.

We further investigate the influence of the number of reverse sampling steps on the proposed diffusion-based channel estimator. This experiment is conducted under the \texttt{TDLB100-400} channel scenario with DMRS configuration type 1 and 3 DMRS symbols per slot. The proposed noise-adaptive posterior correction is applied in all compared cases, and only the number of reverse sampling steps is varied. The specific channel testing scenarios are listed in Table~\ref{tab:ablation_study_step}.

\begin{table}[H]
\caption{Configurations for the ablation study on reverse sampling steps.}
\centering
\setlength{\tabcolsep}{8pt}
\renewcommand{\arraystretch}{1.15}

\begin{tabular}{|>{\columncolor{gray!20}}c|c|c|c|}
\hline
\rowcolor{gray!35}
\textbf{Profile} & \textbf{RMS (ns)} & \textbf{Doppler (Hz)} & \textbf{Steps} \\
\hline
\textbf{TDLB100-400}   & 100  & 400   & 1000/500/200\\ \hline
\end{tabular}

\label{tab:ablation_study_step}
\end{table}

\begin{figure}[H]
\vspace{-3pt}
\centering
\includegraphics[
    width=0.85\columnwidth,
    trim={0mm 55mm 0mm 55mm},
    clip
]{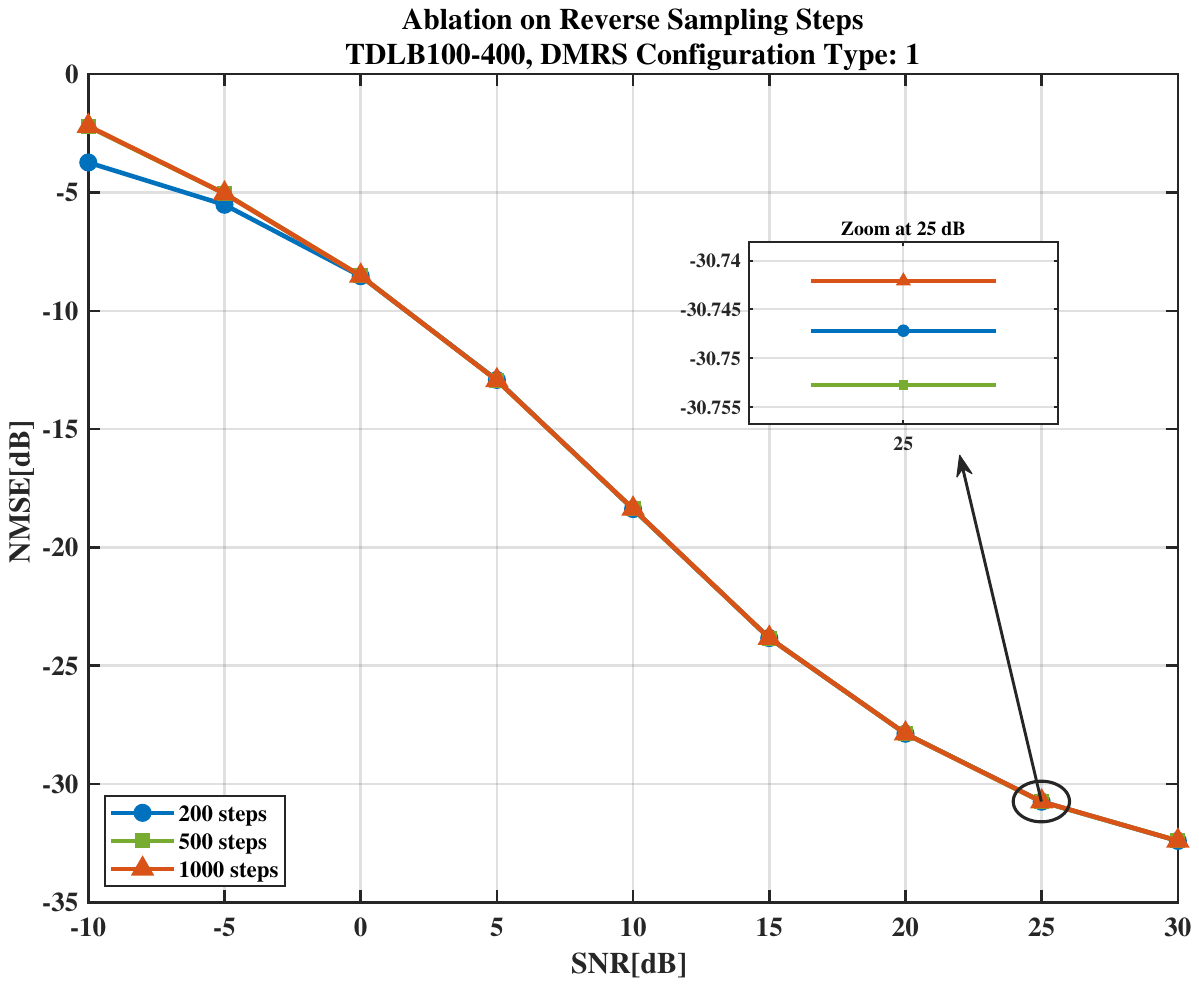}
\caption{Ablation study on the number of reverse sampling steps under the \texttt{TDLB100-400} channel scenario with DMRS configuration type 1 and 3 DMRS symbols per slot. All curves are obtained with the proposed noise-adaptive posterior correction.}
\vspace{-6pt}
\label{fig:sampling_steps_ablation}
\end{figure}

The corresponding NMSE results are shown in Fig.~\ref{fig:sampling_steps_ablation}. It can be observed that the curves obtained with 200, 500, and 1000 sampling steps are highly close to each other over most SNR values. In particular, for SNRs not lower than 0 dB, the performance difference among different sampling steps becomes almost negligible. 

This result indicates that the proposed estimator does not require a very large number of reverse sampling steps to achieve stable reconstruction performance. The reason is that the DDIM-style reverse update provides an efficient sampling trajectory, while the proposed noise-adaptive posterior correction further stabilizes the reconstruction process by regulating the residual stochasticity. Therefore, reducing the number of sampling steps can significantly improve inference efficiency without sacrificing channel estimation accuracy. This finding justifies the use of 200 reverse sampling steps as the default inference setting, which provides a favorable balance between estimation performance and computational complexity.

\IEEEpubidadjcol

\section{Conclusion}
In this paper, we proposed DANCE, a diffusion-based noise-adaptive null-space channel estimation framework for DMRS-aided OFDM systems with sparse and noisy pilot observations. By modeling the received DMRS samples as a sparse linear inverse problem, DANCE exploits the corresponding range--null space structure to combine measurement information from the observed pilot REs with a learned diffusion prior for the unobserved channel components. Unlike direct projection-based methods that may inject pilot noise into the reconstruction, DANCE introduces a noise-adaptive posterior correction mechanism to adjust the range-space correction strength and residual sampling variance according to the observation noise level, thereby balancing pilot consistency and noise suppression during reverse diffusion sampling. An OFDM-tailored conditional U-Net denoiser with classifier-free guidance is further adopted to model complex-valued channel grids under different channel scenarios. Simulation results based on 5G NR TDL and CDL channel models show that DANCE achieves robust full-grid CSI reconstruction under different SNRs, DMRS configurations, maximum Doppler frequency shifts, and train--test distribution mismatches, outperforming practical conventional estimators and diffusion-based posterior sampling baselines in the considered sparse-pilot settings.
\bibliographystyle{IEEEtran}
\bibliography{refs}  

\end{document}